\documentclass[preprint,12pt]{elsarticle}

\usepackage{setspace}
 \usepackage{graphicx}

\usepackage{amssymb}

 \usepackage{amsthm}
\usepackage{fullpage}

\newtheorem{definition}{Definition}

\journal{Physica A}

\begin{document}

\begin{frontmatter}

\title{Assessment of long-range correlation in animal behaviour time series: the temporal pattern of locomotor activity of Japanese quail ({\sl Coturnix coturnix}) and mosquito larva ({\sl Culex quinquefasciatus}).}

\author[JK]{Jackelyn M. Kembro}
\ead{jkembro@efn.uncor.edu}
 \author[GF]{Ana Georgina Flesia}
\ead{flesia@famaf.unc.edu.ar}
  \author[RG]{Raquel M. Gleiser}
  \ead{rgleiser@crean.agro.uncor.edu.}
  \author[JK]{ Mar\'{\i}a A. Perillo}
   \ead{mperillo@efn.uncor.edu}
  \author[JK]{Ra\'{u}l H. Mar\'{\i}n}
  \ead{rmarin@efn.uncor.edu}

\address[JK]{Instituto de Investigaciones Biol\'ogicas y Tecnol\'ogicas (IIByT-CONICET), Instituto de Ciencia y Tecnolog\'ia de los Alimentos, C\'atedra de Qu\'\i mica Biol\'ogica, Facultad de Ciencias Exactas, F\'\i sicas y Naturales, Universidad Nacional de C\'ordoba, 1611 V\'elez Sarsfield, X5016GCA, C\'ordoba, Argentina.}

\address[GF]{Facultad de Matem\'atica, Astronom\'\i a y F\'{\i}sica, Universidad Nacional de C\'ordoba, y Conicet at UTN-Regional C\'ordoba. Av. Medina Allende s/n, Ciudad Universitaria, C\'ordoba.}
\address[RG]{Centro de Relevamiento y Evaluaci\'on de Recursos Agr\'\i colas y Naturales, CONICET. Universidad Nacional de C\'ordoba. Av. Valpara\'\i so  s.n. X5016GCA, C\'ordoba, Argentina. }

\begin{abstract}The aim of this study was to evaluate the performance of a classical method of fractal analysis, Detrended Fluctuation Analysis (DFA), in the analysis of the dynamics of animal behavior time series. In order to correctly use DFA to assess the presence of long-range correlation, previous authors using statistical model systems have stated that different aspects should be taken into account such as: 1) the establishment by hypothesis testing of the absence of short term correlation, 2) an accurate estimation of a straight line in the log-log plot of the fluctuation function, 3) the elimination of artificial crossovers in the fluctuation function, and 4) the length of the time series. Taking into consideration these factors, herein we evaluated the presence of long-range correlation in the temporal pattern of locomotor activity of Japanese quail ({\sl Coturnix coturnix}) and mosquito larva ({\sl Culex quinquefasciatus}). In our study, modeling the data with the general ARFIMA model, we rejected the hypothesis of short range correlations (d=0) in all cases. We also observed that DFA was able to distinguish between the artificial crossover observed in the temporal pattern of locomotion of Japanese quail, and the crossovers in the correlation behavior observed in mosquito larvae locomotion. Although the test duration can slightly influence the parameter estimation, no qualitative differences were observed between different test durations.

\end{abstract}

\begin{keyword}
Detrended Fluctuation Analysis; Fractal; Animal behavior; Crossovers; Trends

\end{keyword}

\end{frontmatter}







\section{Introduction}
\label{intro}

Fractal analysis, such as Detrended Fluctuation Analysis (DFA), has successfully been applied in biology to such diverse fields of interest such as DNA [1], heart rate dynamics [2, 3], neuronal discharges [4], human gait [2, 5, 6], and animal behavior [7-12]. In particular, fractal analysis has been used to evaluate the temporal dynamics of animal behavior in a wide variety of situations and species, ranging from the swimming patterns of Copepoda [13] to social behavior of chimpanzees [7] and even human walking patterns [2, 5, 6]. Various animal behaviors have been shown to present long-range correlations (i.e. for large time lags, the autocorrelation function of such long-memory processes decays according to a power-law and hence exhibits scaling with a characteristic scaling exponent [14], see below). In this context, DFA has emerged as an effective tool to measure the temporal organizational complexity of animal behavior [15, 16], due mainly to the facts that it could be applied to non stationary time series and because it is able to eliminate trends in data. For example, DFA has been found useful to detect subtle changes in behavioral patterns due to sublethal doses of the toxic compound [17], or stressful events [18-20] that could go undetected by conventional behavioral analyses which are often limited to measures of the mean duration or frequency of particular behaviors [10].

Although DFA is used to evaluate temporal patterns of animal behavior, there are some important considerations that must be taken into account for the correct interpretation of results. It is crucial to understand the intrinsic dynamics of the system under study when applying DFA. First, it should be considered that one of the common challenges is the existence of crossovers in the fluctuation function (i.e., a change in the value of $\alpha$ for different ranges of scales) [4, 21]. A crossover may arise from actual changes in the correlation properties of the signal at different time scales (i.e., two levels of complexity), or from trends (a smooth and monotonic or slowly oscillating pattern caused by external effects) that were not correctly eliminated by the DFA [22, 23]. The existence of trends in time series generated by biological systems is very common and almost unavoidable [22]. For this reason, efforts should be made to systematically study the trends in time series data (for complete review see Hu et al. [22]; Kantelhardt et al. [23]). However, most of the biological studies that use DFA do not appear to consider the presence of trends of orders higher than one in their time series.

Second, Maraun et al. [14] proposed that when using DFA, long-memory should not be assumed a priori but must be established. To reliably infer power-law scaling of the fluctuation function, a straight line in the log-log plot has to be established. This requires the investigation of the local slopes [14]. However, finite datasets bring along natural variability. To decide if a fluctuating estimation of the slope should be considered as being constant, empirical confidence intervals must be calculated for a long-range and a simple short-range correlated model [14]. To the authors knowledge, this has never been applied to animal behavior time series.

Third, the duration of the test and sampling frequency are of vital importance in animal behavioral studies. Many animal behavioral studies last only five or ten minutes. Longer durations are not always possible because behavioral pattern and experimental conditions can change through time. For example, behavioral studies often focus on studying animals in novel, stressful environments where the animal can adapt over time, or they evaluate response to a drug or toxin which metabolizes in the organism resulting in the blood concentration changing over time. In addition, animal behavior is subject to circadian rhythms, and exhaustion if forced to perform a certain behavior over a prolonged period of time. The sampling frequency of the behavior also has limitations. In the case of evaluating locomotor activity, the sampling interval cannot be shorter than the time it takes the animal to take a step. Hence, there are empirical limitations on the number of data points a behavioral time series can have. Studies that use DFA to evaluated changes in the correlation properties of a behavior due to stressors or age, many times have around 1400 - 7200 data points [5, 6, 9, 10, 12, 17, 19, 20], for example representing test durations no longer than 1 h, with sampling intervals between 0.3 to 1 s. The length of the time series is also important to accurate estimate the scaling parameters and in evaluating whether the process presents short-range or long-range correlation. In a short memory processes, the slope of the fluctuation function converges to $\alpha =0.5$ for time scales large enough. However, for a finite set of data, a priori one cannot be sure that the series is long enough to show this plateau. Therefore, for a process with unknown correlation structure, it is misleading to use $\alpha >0.5$ as evidence for long-memory (see Section 3.2 for further details) [14], especially when the time series is short. In other words, it might be possible that the record is too short to exhibit a plateau with $\alpha=0.5$. Herein, the effect of the duration of the behavioral time series on estimation of scaling parameters will be evaluated.

 Forth, in a previous study [18-20] in Japanese quail, we have shown that long periods of inactivity (relative to the total duration of the test) caused by a heighten fear response can result in a loss of the typical monofractal pattern. Thus, it is important to consider this factor when evaluating behavioral time series with DFA.

In this paper, we study the use of fractal analysis to evaluate the temporal dynamics of the locomotor pattern in two very different animal models, Japanese quail ({\sl Coturnix coturnix}) and larvae of {\sl Culex quinquefasciatus} mosquito. The Japanese quail is a bird that principally walks, although in certain stressful conditions can make short flights. On the contrary, the larvae of {\sl Culex quinquefasciatus} mosquito dwell in the water, feeding on organic material in the water, and must come to the surface to breath. The paper is organized as follows: In the next section the experimental details are described. In Section 3, first, we investigate the slopes in the fluctuation function and determine empirical confidence intervals for a family of correlated models, the ARFIMA(p,d,q) models. In Section 4, first we systematically study different orders of the DFA technique to assess the presence of trends in the data time series. Second, we investigate the local slopes in the fluctuation function of DFA. Third, we evaluate the effect of test duration on the estimation of the DFA self-similarity parameter, $\alpha$. Section 5, compares DFA with a different fractal analysis that has been used in animal behavior studies, Frequency distribution of behavioral events or symbolic analysis [13, 24-26]. We summarize the results in Section 6.

\section{Experimental setting and data recording}

Japanese quail were reared following Kembro et al. [27]. At 31 days of age, 4 female and 4 male quail were individually housed in a white wooden box measuring 43 x 41.7 x 46 cm (length x width x height). Each box had a sand floor, a wire-mesh roof (5 cm grid) and a video camera placed 1.3 m above the box. Coincident with placement into their boxes, birds were switched to a quail breeder ration (20\% CP; 2900 Kcal ME/kg) and water was continued ad libitum. Quail were subjected to a daily cycle of 14 hours light (between 6:00 and 20:00):10 hours dark during the study. The next day (32 days of age) the animal's behavior was recorded onto computer during 12 consecutive hours, from 07.00 to 19.00 hours. During this observation period, the experimenter did not enter the room.

Forth instar {\sl Culex quinquefasciatus} larvae were reared in the laboratory (following [28]). Seven larvae were individually and simultaneously tested in identical experimental glass boxes (4 cm x 4 cm x 16 cm) with one transparent and 3 white vertical walls. The seven boxes were lined approximately 3 cm apart, so that each larva could not visualize the larvae in the next neighboring box. These seven larvae were assigned to one of two treatments that differed in the type of water used during the experiment, distilled water or source (water from the container the larvae were raised in). The assay was recorded during 3 hours with a video camera installed directly in front of the boxes. The test duration was chosen in order to minimize the effect that a lack of food supply could have on locomotor activity. During this observation period, the experimenter remained out of the room to minimize disturbances.

The locomotor activity of the animal was determined from the video recordings using the ANYMAZE computer program. The X, Y coordinates of the animal were recorded at 1/2 second (quail) or 1/3 second time interval (mosquito larvae). A value of 1 represented the animal moving during the time interval (and the distance ambulated was also recorded), or alternatively a 0, representing immobility, was assigned if the animal did not move during the time interval. Thus, a time series ($x_i$) of the locomotor activity during the test was constructed for each animal, with a total $N$ of 86400 or 32400 data points, quail or mosquito larvae respectively. The following variables were measured for each animal:

\begin{itemize}

\item {\sl Percentage of total time ambulating:	}	
\[
t{\%}=\frac{\sum t_i}{N} . 100
\]
where $t_i$ is the time interval (s) in which the animal is ambulating and $N$ is the total number of data points of the test (s).
\item {\sl Ambulation event:} interval of time ($ > 0.6$ s) in which the animal moves continuously.
\item 	{\sl Immobility event:} interval of time ($ > 0.6$ s) in which the animal remains immobile.
\end{itemize}

\section{Definition and Properties of Detrended Fluctuation Analysis (DFA) Method}
\subsection{Definition}

Long-range correlated processes are characterized by algebraically decaying correlations. Given a time series $\{X_t\}$, the Detrended Fluctuation Analysis (DFA) proposed by [1], and described in detail in [18] consists of five steps. In the first one, for each  $t\in\{1,\dots,N\}$, the cumulative walking time series $Y_t$
\begin{equation}
Y_t=\sum_{j=1}^t X_j
\end{equation}
is computed. This integrated time series $\{Y_t\}$ is divided into $[N/n]$ non overlapping blocks, each containing $n$ observations. In the third step, for each block, a least square line is fitted to the data, which represents the local trend of the block. In the fourth step, the time series  $\{Y_t\}$ is detrended:
\begin{equation}
Z_t=Y_t-Y_t^n
\label{eqqq}\end{equation}
 where $Y_t^n$ denotes the adjusted fit on each block.


Finally, in the fifth step, for each $n\in \{2m+2,\dots,N/4\}$, the root mean square fluctuation function $F(n)$ is computed.
\begin{definition}
The root mean square fluctuation function $F(n)$ is defined by
\begin{equation}
F(n)=\sqrt{\frac{1}{M}\sum_{t=1}^{M} (Y_t-Y_t^n)^2}=\sqrt{\frac{1}{M}\sum_{t=1}^{M} Z_t^2}
\end{equation}
where $M$ is maximum multiple of $n$, smaller or equal to $N$, i.e. $M=[N/n]$.
\end{definition}

Observe that $F(n)$ will increase with block size $n$.  A linear relationship on a log-log scale indicate the presence of power law scaling
\begin{equation}
F(n)=\phi n^\alpha
\label{eq4}
\end{equation}
Under such condition the fluctuations can be characterized by a  scaling exponent $\alpha$, which is the slope line when regressing $\ln(F(n))$ on $\log(n)$.

By taking the logarithm of the toot mean square fluctuation value, given by \ref{eq4} we obtain
\begin{equation}
\log(F(n))=\log(\phi)+\alpha\log(n).
\label{eq5}
\end{equation}
Then, we may rewrite eq.( \ref{eq5}) in the context of a simple linear regression model, obtaining an estimate of $\alpha$ given by
\begin{equation}
\hat \alpha=\frac{\sum_{j=1}^m (x_j -\overline x)y_j}{\sum_{j=1}^m (x_j -\overline x)^2}=\frac{\overline x(1-\overline y)}{\frac{1}{m}\sum_{j=1}^m(x_j -\overline x)^2}
\end{equation}
where $y_j$ are $\log(F(n))$, $m=N/4-(2m+2)$ and $x_j=log(n)$.

The $\alpha$-value relates to the autocorrelation structure of the original time series $X_i$ in the sense that if  $\alpha= 0.5$ indicates that the original series is uncorrelated (random). Short range correlations (correlations decay exponentially) would be indicated if the log-log plot approached a straight line with slope 0.5 for large window sizes. The situation of $0.5 <\alpha   < 1$ indicates long-range autocorrelation (correlations decaying as a power-law) exist, meaning that on-going behavior is influenced by what has occurred in the past, [18, 30].

\subsection{Properties}
Different properties have been established by simulation when the true process is a fractional Brownian process [29], and  ARFIMA(p,d,q) processes [30], even in the presence of trends. However, the length of the series considered on these papers are short enough to generate doubts about the estimation accuracy of  $\alpha$, when working with real data.  Maraun et. al. [14] have given warnings about misleading estimation with data simulated with ARMA processes with superimposed trends, when the series are too short to see the plateau a = 0.5.
In our case, we have series long enough to consider valid Crato et al. [30] asymptotic
estimations, which we are introducing below for the sake of completeness.

\subsubsection{Asymptotic normality}
If the $F(n)$ function is contaminated by an additive independent gaussian noise with standard deviation $\sigma$, the  $\hat \alpha$ estimator obtained by Ordinary Least Squares (OLS) is a consistent estimator with variance given by
\begin{equation}
Var(\hat \alpha)=\frac{\sum_{j=1}^m (x_j -\overline x)^2 Var(y_j)}{\left(\sum_{j=1}^m (x_j -\overline x)^2\right)}=\frac{\sigma^2 }{\frac{1}{m}\sum_{j=1}^m(x_j -\overline x)^2}
\end{equation}
The distribution is asymptotically normal, so a hypothesis test for the presence of long term correlation can be expressed as

\[H_0: \alpha=0.5 \quad H_A: \alpha\neq 0.5\]

\noindent In fact, in the case of ARFIMA(p,d,q) processes, the differentiating parameter $d$ is related to the $\alpha$ parameter, since $\alpha=d+0.5$, so the test based on DFA scaling parameter is usually written as
\[H_0: d=0 \quad H_A: d\neq 0\]
The test statistics for the DFA estimator $\hat d=\hat \alpha-1/2$ is given by
\[
Z=\frac{\hat d-d_{H_0}}{\sqrt{Var(\hat d)}}.
\]
The lower and upper confidence interval values for the parameter $d$ based on any of the estimation methods proposed here, are given by

lower value = $\hat d - z_{\alpha/2}. \sigma_{\hat d}$  \hspace{3cm}  upper value = $\hat d + z_{\alpha/2}. \sigma_{\hat d}$

\noindent where $z_{\alpha/2}=1.96$ and $ \sigma_{\hat d}=\sqrt{Var(\hat d)}$.
\subsubsection{Detrending order}

The method of Detrended Fluctuation Analysis is an improvement of classical Fluctuation Analysis, which is similar to Hurst's rescaled range (R/S) analysis. They allow the empirical determination of the correlation properties on large scales. All three methods are based on random walk theory, but while both, FA and Hurst's methods fail to determine correlation properties if linear or higher order trends are present on data, DFA explicitly deals with the monotonous trends in a detrending procedure. This is done by estimating a piecewise polynomial trend $y(p)(s)$ within each segment by least square fitting. Conventionally the DFA is named after the order of the fitting polynomial (DFA1, DFA2, etc) Note that DFA1 is equivalent to Hurst's analysis in terms of detrending.

Bashan et. al. [31] remarked that only a comparison of DFA results using different detrending polynomials yields full recognition of the trends. Also, a comparison with independent methods is recommended for proving long range correlations.

In the next section, we discuss the test for long range correlation under the ARFIMA models for two different decorrelation orders of DFA, linear and cubic. In one case, linear detrending is enough for establishing long term correlation, but in the other, linear and cubic detrending yield significantly different estimations, showing the need for a deeper study of trends.

\subsection{Behavioral time series}
In this subsection we analyze times series related to 8 different Japanese quail and 7 mosquito larvae. For each sequence we tested the long memory hypothesis
\[H_0: d=0 \quad H_A: d\neq 0\]
under the ARFIMA(p,d,q) general model.

For each sequence we represent graphically the 95\% confidence intervals for the long memory parameter $\alpha=d+1/2$, with $d$ the fractional parameter of the allegedly ARFIMA model.

\subsubsection{Japanese Quail {\sl Coturnix coturnix}}
In Table 1 we observe that the existence of a long range dependence is statistically significant at a 5\% level for the DFA estimation of the differential parameter $d$ in the 8 quail behavioral time series. Taken into consideration the confidence intervals for the different animals studied, the short memory hypothesis ($\alpha = 0.5$) can be rejected with good power. The table also shows the confidence intervals for two different DFA detrending procedures, DFA1 and DFA3. Although for some animals (3, 4 and 5) the confidence interval was numerically larger, overall the difference in the detrending order is not significant, since in all cases, the confidence intervals cross over.
\begin{table}[h]
\centering
\begin{tabular}{|c|c|c|}
  \hline
  DFA1 $\pm$ SE*1.96 & DFA3 $\pm$ SE*1.96\\ \hline
 $ 0.7875     \pm 0.0125$ & $0.8061\pm  0.0155$\\
 $ 0.7727     \pm 0.0149 $ & $0.7726\pm 0.0153$ \\
 $  0.7844    \pm 0.0186 $ & $ 0.7935\pm 0.0147$ \\
  $ 0.7907   \pm 0.0259 $ & $ 0.8037 \pm 0.0153$ \\
  $0.7722    \pm 0.0196 $ & $0.8054\pm 0.0149$ \\
  $0.8090     \pm 0.0123 $ & $0.8089\pm 0.0178$ \\
  $0.8956    \pm 0.0172 $ & $0.8838 \pm 0.0196$ \\
  $0.8131    \pm  0.0202 $ & $0.8116 \pm 0.0198$ \\
  \hline
\end{tabular}
\caption{Confidence intervals for the $\alpha$-value of behavioral time series of Japanese quail estimated with a DFA detrending order of 1 (DFA1) and 3 (DFA3) and at a 95\% confidence level.}
\label{tablacodorniz}
    \end{table}

\subsubsection{Mosquito larva {\sl Culex quinquefasciatus}}
From Table 2 we observe that the existence of a long range dependence is statistically significant at 5\% level for the DFA estimation of the differential parameter $d$, in the 7 behavioral time series of mosquito larvae evaluated.

Taking into consideration the confidence intervals estimated for the different animals the short memory hypothesis ($\alpha= 0.5$) can be rejected. The power of the test in this case is not as good as in the case of the Japanese quail, showing a much larger confidence interval in mosquito larvae (Table 2; range for DFA3: 0.0195-0.0343) than that observed for Japanese quail (Table 1; range for DFA3: 0.0147- 0.0198). These differences between species will be further explored in the following sections. Table 2 also shows the confidence intervals estimated for two different DFA detrending procedures, DFA1 and DFA3. Significant difference in the $\alpha$-value estimated with DFA1 and DFA3 was observed in the behavioral time series from animal 3, 5 and 7, where the confidence intervals cross each other. The differences between detrending orders imply the presence of crossovers (see Section 4), and generate the need of a deeper study of the scaling properties of the behavioral series of this species.

\begin{table}[h]
\centering
\begin{tabular}{|c|c|c|}
\hline
DFA1 $\pm$ SE*1.96 & DFA3 $\pm$ SE*1.96\\
  \hline
 $ 0.6651 \pm 0.0264$ & $0.7078\pm  0.0301$\\
 $ 0.6654 \pm 0.0242 $ & $0.670\pm 0.0277$ \\
 $ 0.7287\pm 0.0357 $ & $ 0.8005\pm 0.0343$ \\
  $0.6201\pm 0.0184 $ & $0.6152 \pm 0.0195$ \\
  $0.7163\pm 0.0384 $ & $0.8028\pm 0.0263$ \\
  $0.5450 \pm 0.0227 $ & $0.5707\pm 0.0248$ \\
  $0.8034\pm 0.0324 $ & $0.8939 \pm 0.0235$ \\
  \hline
\end{tabular}
\caption{Confidence intervals for the $\alpha$-value of behavioral time series of mosquito larvae estimated with a DFA detrending order of 1 (DFA1) and 3 (DFA3) and at a 95\% confidence level.}
\label{tablamosquito}
    \end{table}

\subsubsection{Word of caution}
Maraun et al. [14] and Kantelhardt et al.  [23] have given warnings about the abuse of such asymptotic results without proper model estimation. Series length and high degree trends may reduce the power and size of such estimations, so it is necessary to study several degrees of detrending to determine that the model is not indeed a masked short term correlation process, and to achieve stationarity.

Crato. et al. [30] have also shown a real example with four DNA sequences of more than 300000 bp each. All of them reject the null hypothesis of short memory with d values as small as 0.032, if the ARFIMA model is valid. Rejections of 10 different estimators validate the results. In the following sections we will study several order of detrending within the DFA procedure, and the weight of the series length in the DFA analysis.

\section{Detrending}
Detrending can be defined as the operation of removing the trend; whereas a trend is an intrinsically fitted monotonic function or a function in which there can be at most one extremum within a given data span [32]. Kantelhardt et al. [23] reported that trends in the original time series data can lead to an artificial crossover in the slope of the log-log plot of $F(n)$ vs. n, i.e., the slope $\alpha$ is increased for large time scales. To determine the order of DFA that would eliminate these trends, and to estimate the value of $\alpha$ reliably,reliably, DFA was calculated with different detrending orders. For this purpose, linear (DFA1), square (DFA2), cubic (DFA3) and higher order polynomials were used in the fitting procedure. Since the detrending of $Y_j$ time series is performed by subtraction of the fits from data, these methods differ in their capability of eliminating trends in the data. In the mth-order DFA (DFAm), trends in the profile of the $Y_j$ time series of order m-1 are eliminated [23]. In other words, the artificial crossover (more than one linear fit) disappears when the detrending order used in the DFA is larger than the order of the trend. For example, if linear trends are present in the original time series, DFA2 and higher would eliminate the artificial crossover produced by linear trends. Therefore, after evaluating the fits of all DFA performed with different detrending orders, we selected the lowest detrending order that eliminated trends in all data series and used this order for group comparisons.

\subsection{Elimination of trends in data}

As previously stated, for the reliable detection of long range correlations, it is essential to distinguish trends from the long-range fluctuations intrinsic in the data. It is the advantage of the DFA that it can systematically eliminate trends of different order. In this way, we can gain an insight into the scaling behavior of the natural variability as well as into the trends in the considered time series [23].

First, a Power Spectrum Analysis was used to evaluate the presence of can regular oscillations in time-series data. This is important given that a superposed sinusoidal trend can cause artificial crossover in the fluctuation function at the scale corresponding to the period of the sinusoidal trend [22]. For this reason, in order to accurately use DFA, a given time series should not present periodic oscillations. Thus, their presence were evaluated using the power spectrum analysis tool (POWSPEC.XFM) from SigmaPlot [33]. No characteristic periodic oscillations were observed in any of the time series, indicating that DFA can appropriately and directly be used herein (data not shown).

The possible presence of trends in the locomotor data time series of quail and mosquito larvae (Figure  1a,b, and 2a,b, respectively) was systematically studied by applying consequently higher polynomial orders to the fitting procedure of DFA (see previous section). Several locomotor time series of Japanese quail showed more than one slope (artificial crossovers) when using DFA1 and DFA2 (Figure  1c). However, these crossovers disappeared when DFA of order 3 was used, thus showing that the presence of trends in the data have led to an artificial crossover in the scaling behavior of the fluctuation function, i.e., the slope is increased for large time scales $n$ [23]. Importantly, these crossovers disappear for higher detrending orders. These results are consistent with the analysis shown in Table 1 (Section 3.3.2) were in some animals (3, 4 and 5) the confidence interval for the estimation of $\alpha$ with DFA1 was numerically larger than with DFA3, reflecting more dispersion in the Fluctuation as a result of the presence of crossovers for DFA1. In all, these results further indicate that the DFA method is a reliable tool to accurately quantify correlations [22] in biological signals embedded in polynomial trends.

The position of this artificial crossover depends on strength $A$ and power $p$ of the trend [23]. For weak trends no artificial crossover is observed if the detrending order 1 is larger than $p$ [23]. Thus, the reported scaling and crossover features of $F (n)$ can be used to determine the order of polynomial trends present in data [22].

\begin{figure}[h]
\centering
\includegraphics[width=5cm]{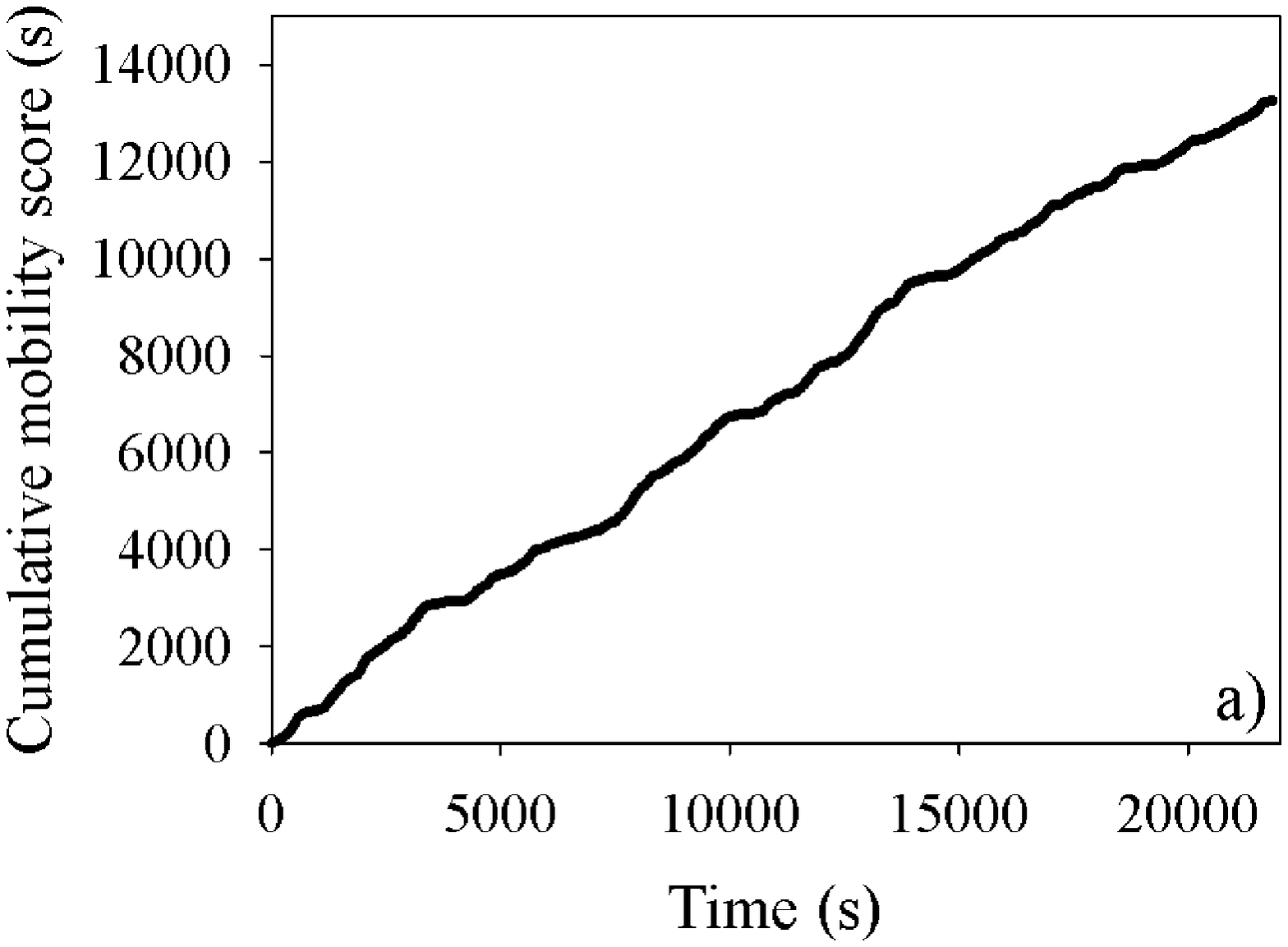}\includegraphics[width=5cm]{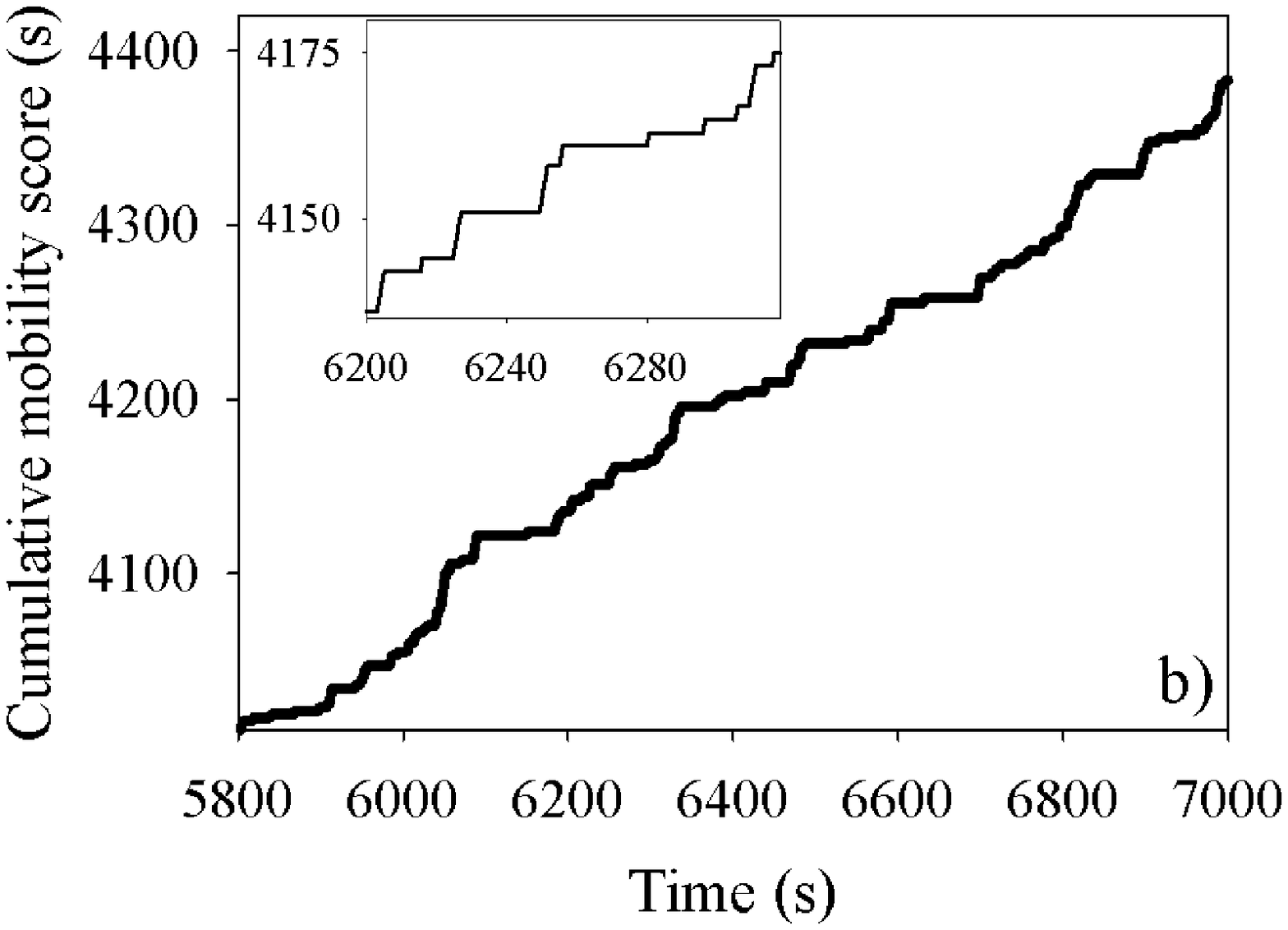}\includegraphics[width=5cm]{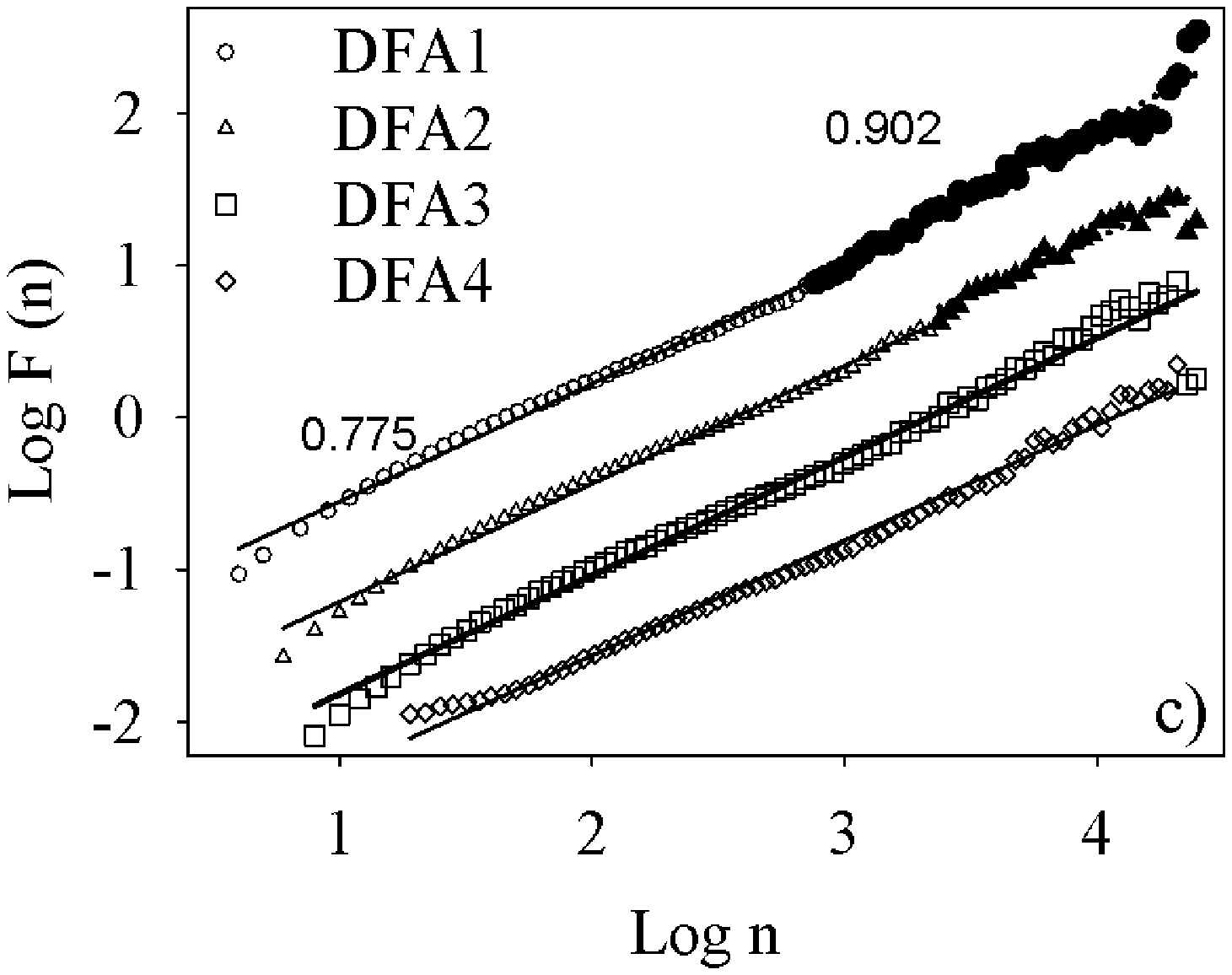}\\
\includegraphics[width=5cm]{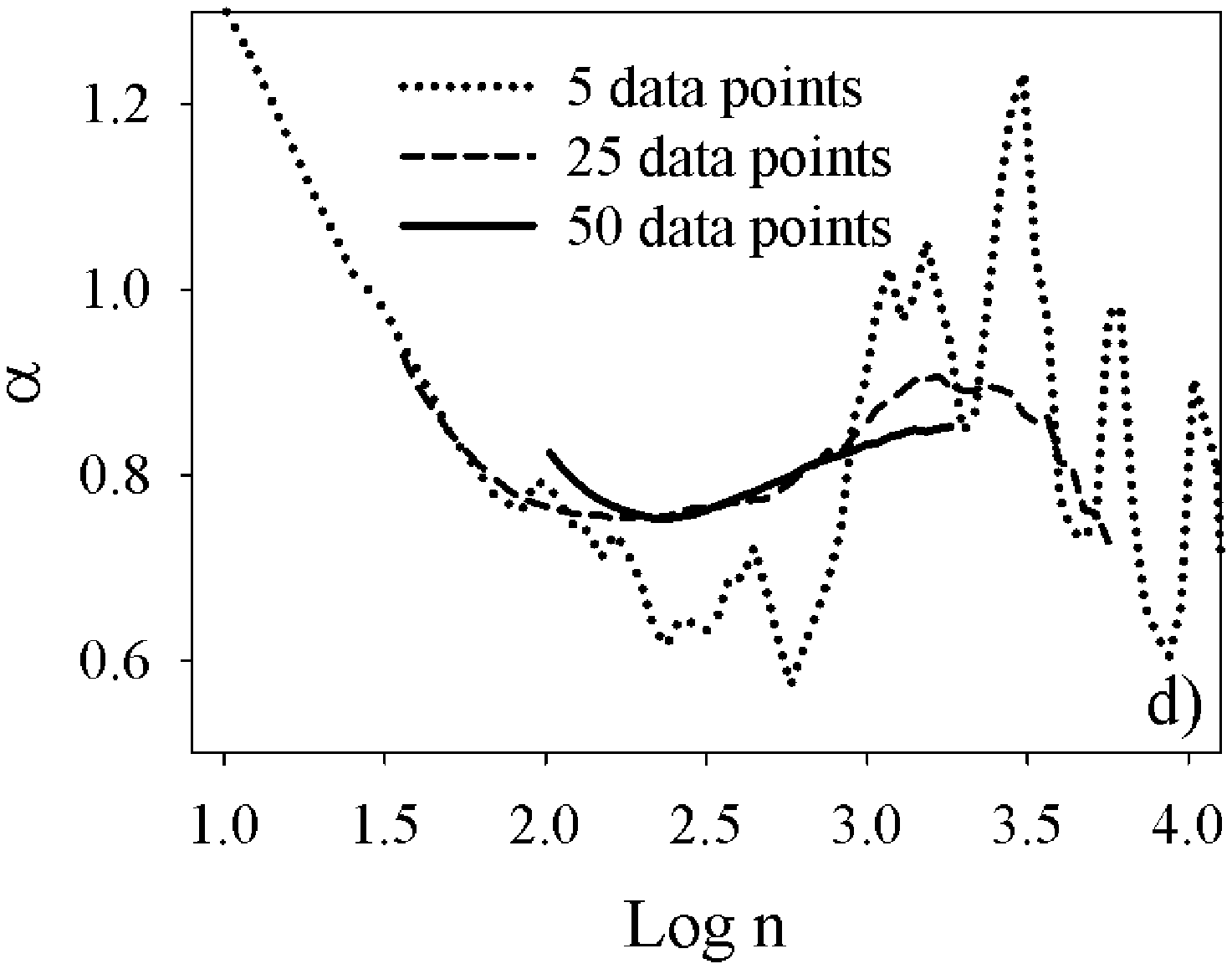}\includegraphics[width=5cm]{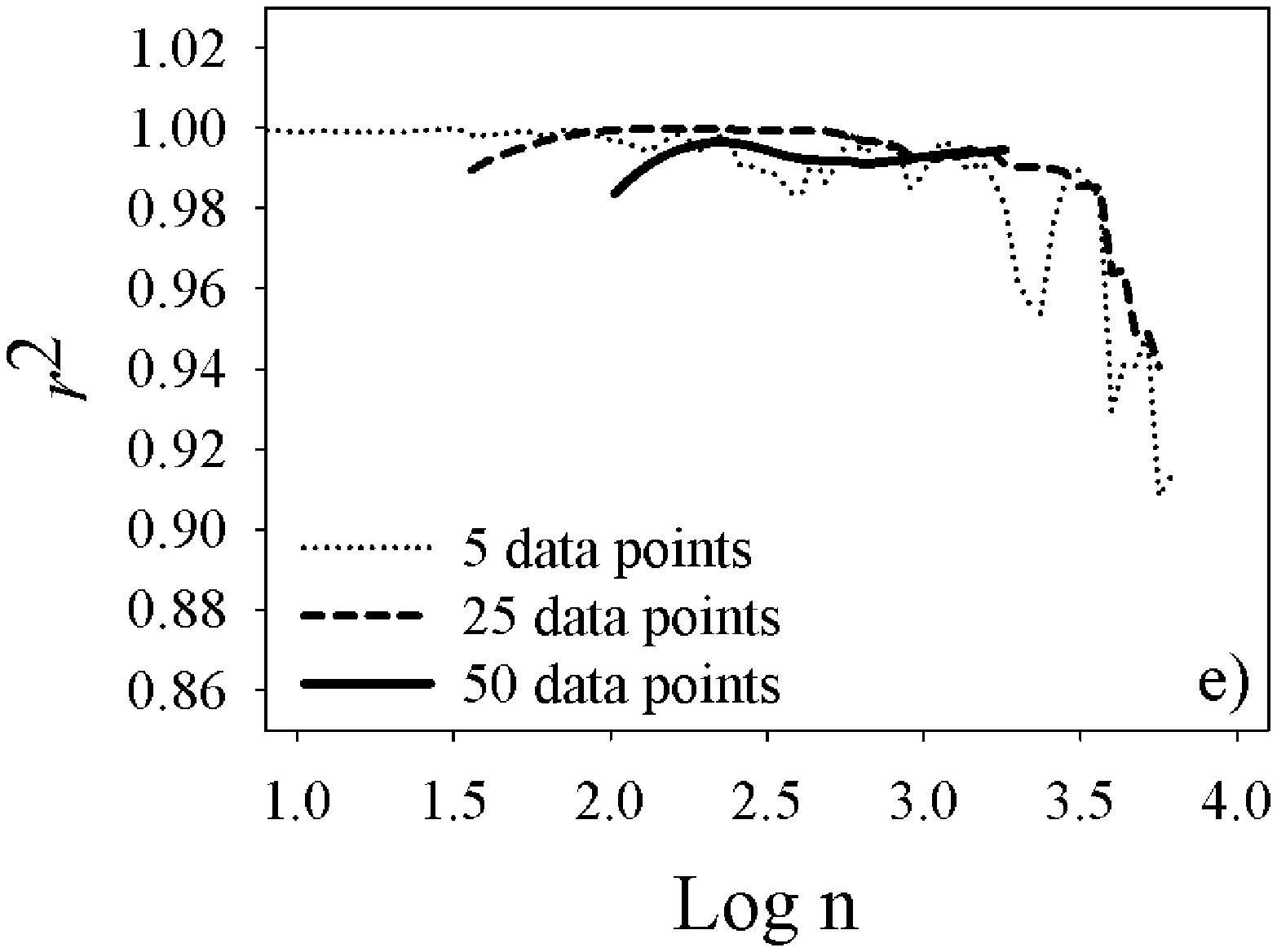}\includegraphics[width=5cm]{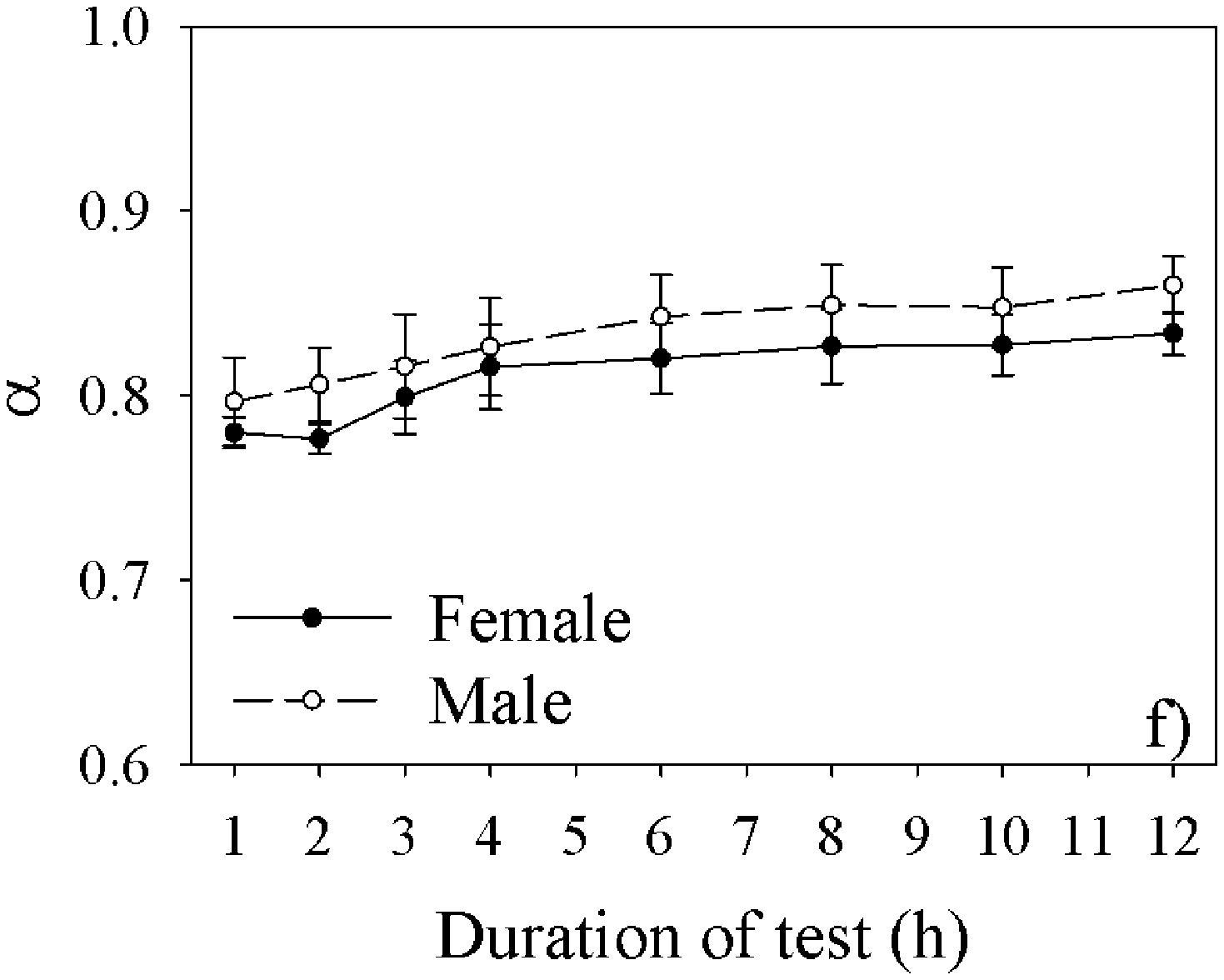}
\caption{. Detrendend Fluctuation Analysis of locomotor temporal pattern of Japanese Quail evaluated in their home cage. a) Cumulative locomotor time series; b) Magnification of the same time series as ``a", a similar pattern in observed for all levels of magnification. c) Example of trend elimination capability of DFA in time series using linear (DFA1), quadratic (DFA2), cubic (DFA3) and 4th order (DFA4) DFA. DFA3 was selected for group comparisons, given that 3rd order polynomials eliminated trends in data. d) Local slopes ($\alpha$) and e) $r^2$ of the corresponding linear fits estimated for different window sizes (5, 25 and 50 data points). Note that for small windows, the bias is very low, but the variability renders the interpretation difficult, whereas for large windows, the variance is reduced at the cost of a biased estimate of $\alpha$  estimated with DFA, and f) for increasing test durations. Values are represented as mean $\pm$ S.E. A one-way repeated measures ANOVA was used to determine the effects gender (female and male), and test duration (within-subject factor) as well as their interactions on the estimation of $\alpha$. No significant effect of gender or test duration was observed.}
\end{figure}

 For the locomotor time series of quail, DFA3 (third order) is the lowest detrending order that eliminated trends in all data time series, and therefore was used in the rest of this study. An ANOVA showed no differences $(P > 0.05)$ between male and female quail in their p-values ($0.86 \pm 0.01$ and $0.84 \pm 0.02$; respectively) nor in their values of $r^2$ of the corresponding linear fit ($0.995 \pm 0.000$ and $0.995 \pm 0.001$; respectively).

For mosquito larvae locomotor time series data, a crossover in the correlation behavior is observed (Figure  2c). The crossover is clearly visible in the results for all detrending orders l. All detrending orders showed almost identical slopes In addition, rather similar crossover positions were observed for all detrending orders, with a slight systematic deviation that is most significant in the DFAl with higher order detrending as predicted by Kantelhardt et al. [23] in their work with artificial time series.

The crossovers were observed for window sizes ranging from $\log n \in[ 2.49 - 3.22]$ (mean value $2.83 \pm 0.13$); which implies that a change in the autocorrelation properties of the locomotor activity of mosquito larva occurred at a temporal scale around 330 seconds (range [ 155 - 830] seconds). The value of the first slope varies slightly between animals $\alpha_1 = 0.97 {\pm} 0.02$ (range [0.93  - 1.09]) with a good linear fit ($r^2 = 0.996 {\pm} 0.001$). On the contrary the value of the second slope is lower and is much more variable between animals ($\alpha_2 = 0.42 {\pm} 0.06$ range [0.18 - 0.73]), also the corresponding linear fit is poorer and more variable ($r^2 = 0.88 {\pm} 0.05$; range [0.65 - 0.97]). If not taken into account these crossovers can significantly impair the estimation of the $\alpha$-value as seen in Section 3.3.3, where a larger 95\% confidence interval was observed for mosquito larvae in comparison to Japanese quail, thus hindering interpretation of results.

The value of the first slope varies slightly between animals $\alpha_1 = 0.97 {\pm} 0.02$ (range [0.93 - 1.09]), with a good linear fit. On the contrary the value of the second slope is lower and is much more variable between animals (also the corresponding linear fit is poorer and more variable).
In all, DFA was able to distinguish between artificial and real crossovers in behavioral time series. The artificial crossover observed in the temporal pattern of locomotion of Japanese quail (Figure  1c) was eliminated for detrending orders 3 and higher; and was clearly different than the real crossovers in the correlation behavior observed in mosquito larvae locomotion, which showed identical slopes and rather similar crossover positions for all detrending orders (Figure  2c). These results highlight the importance of systematically studying the presence of trends in behavioral time series when applying DFA.

\subsection{Establish scaling}

Maraun et al. [14] proposed that to reliably infer power-law scaling of the fluctuation function, a straight line in the log-log plot has to be established. Since a straight line is tantamount to a constant slope, the local slopes $n$ of $\log(F(n))$ vs. $\log(n)$ have to be evaluated for constancy in a sufficient range. The author noted two difficulties for the calculation and interpretation of the local slopes in finite time series: First, estimating the local slopes by finite differences results in a large variability. This can be reduced fitting a straight line to log $F (n)$ vs. $\log n$ within a small window. The window is then shifted successively over all calculated scales $n$. Maraun et al. [14] observed that choosing the optimal window size, one has to trade bias for variance: for small windows, the bias is small, but the variability renders the interpretation difficult, whereas for large windows, the variance is reduced at the cost of a biased estimate of $\alpha$. Thus, the extreme case of a single straight line fit to the whole range of scales considered is maximally biased.

Figure  1d and 2d show the local slopes calculated from the locomotor time series of quail and mosquito larva, respectively. Also, we estimated the $r^2$ of the linear fit for the local slops (Figure  1e and 2e). When applying the DFA method to locomotor time series of Japanese quail (Figure  1c) a slight bend-down is observed for DFA3 for a very small box of $F(n)$, this is because many variables are needed to fit those few points [22]. This is clearly observed in Figure  1c by an increased $\alpha$-value for small box sizes. Also, for very large boxes, fluctuations become larger (Figure  1d) and the $r^2$ of the estimation decreases (Figure  1e), due to the under sampling of $F(n)$ when $n$ gets closer to the length of the signal $N$ [22]. For this reason Hu et al. [22] proposed that one-tenth of the signal length could be considered as the maximum box size when using a DFA, and not $N/4$ that is frequently used.

When applying the DFA method to locomotor time series of mosquito larvae an abrupt decline in the $\alpha$-value (Figure  2d) and the $r^2$ (Figure  2e), of the estimation is observed for increasingly large window, clearly representing a crossover in the fluctuation function.

\begin{figure}[h]
\centering
\includegraphics[width=5cm]{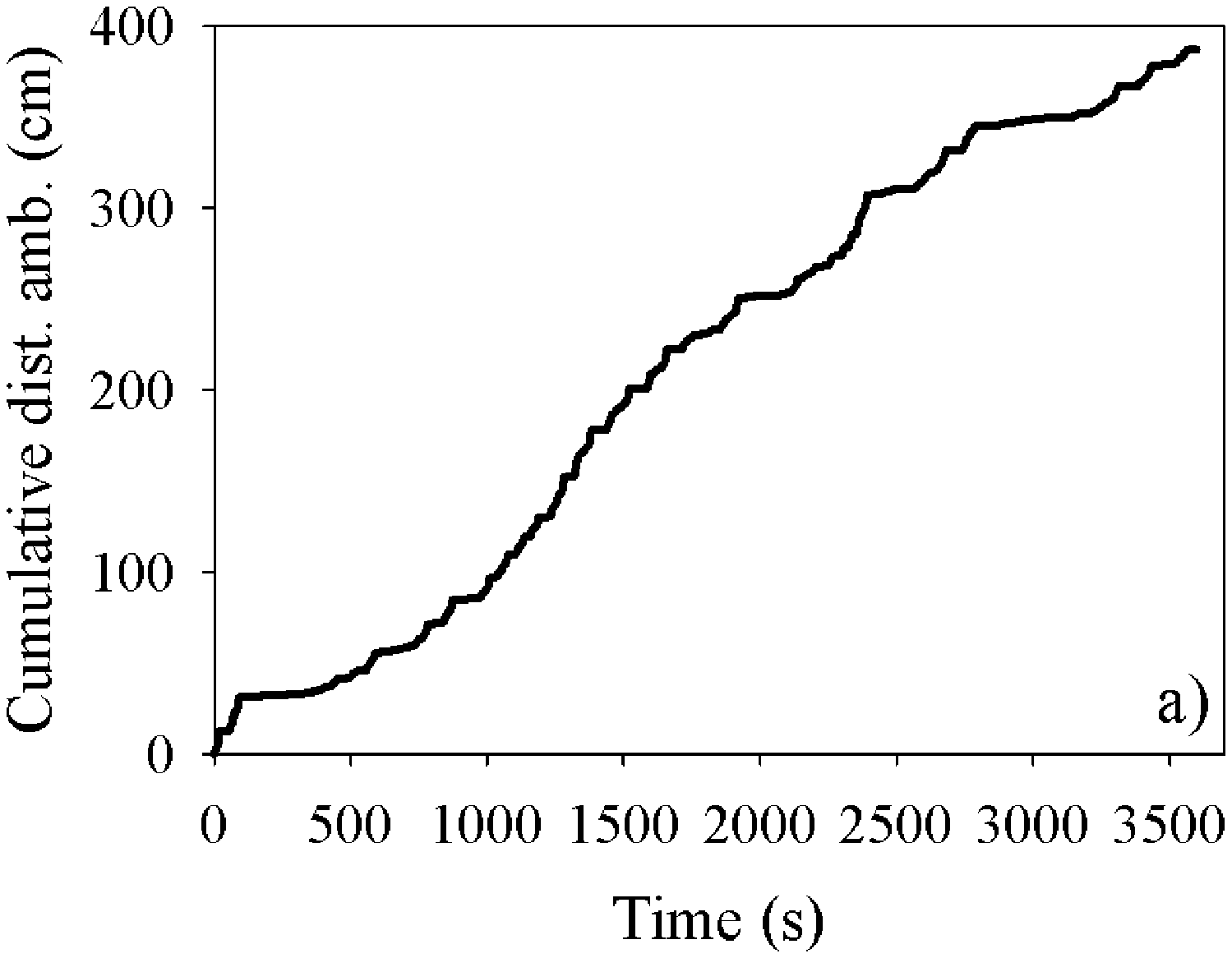}\includegraphics[width=5cm]{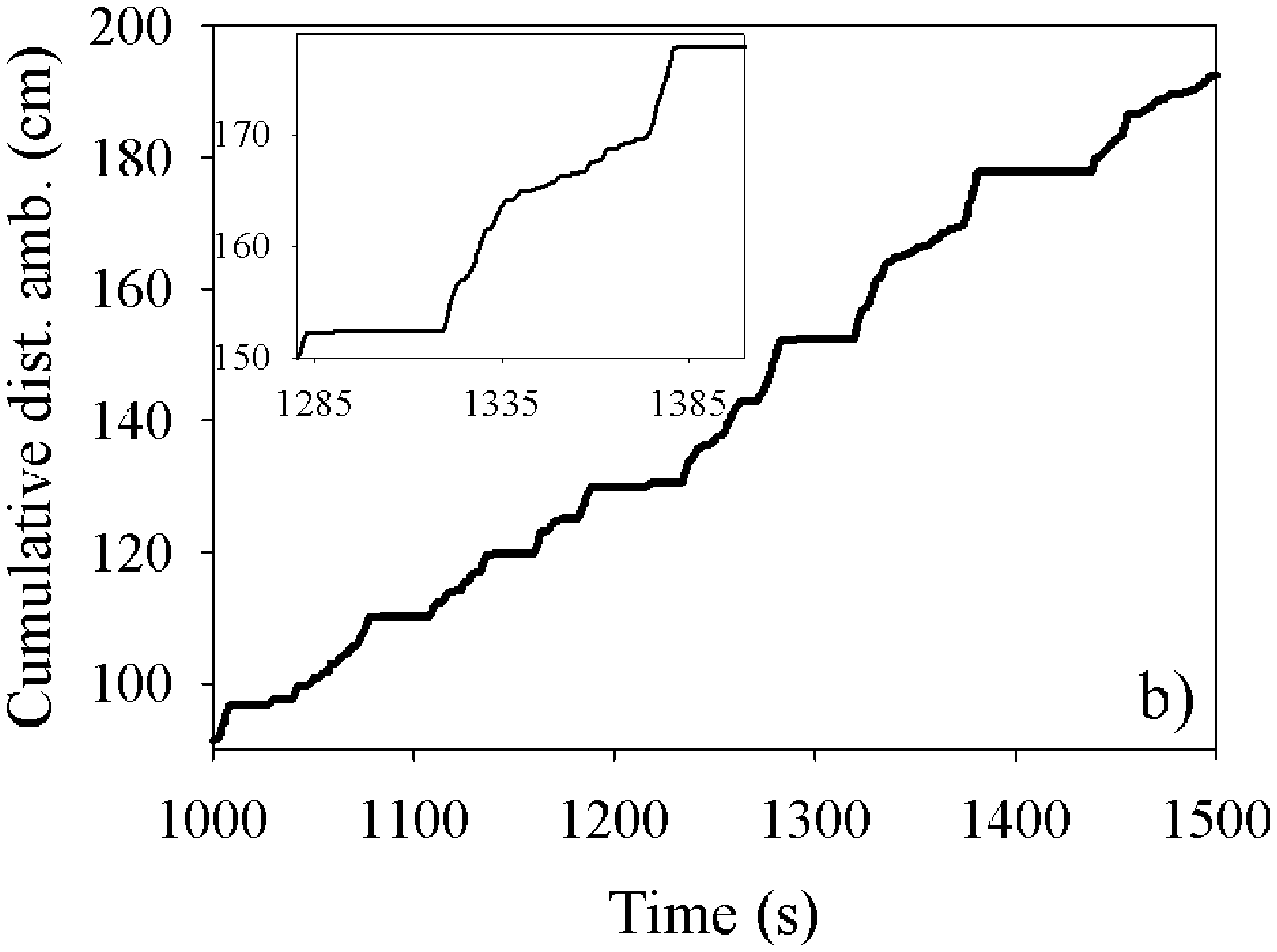}\includegraphics[width=5cm]{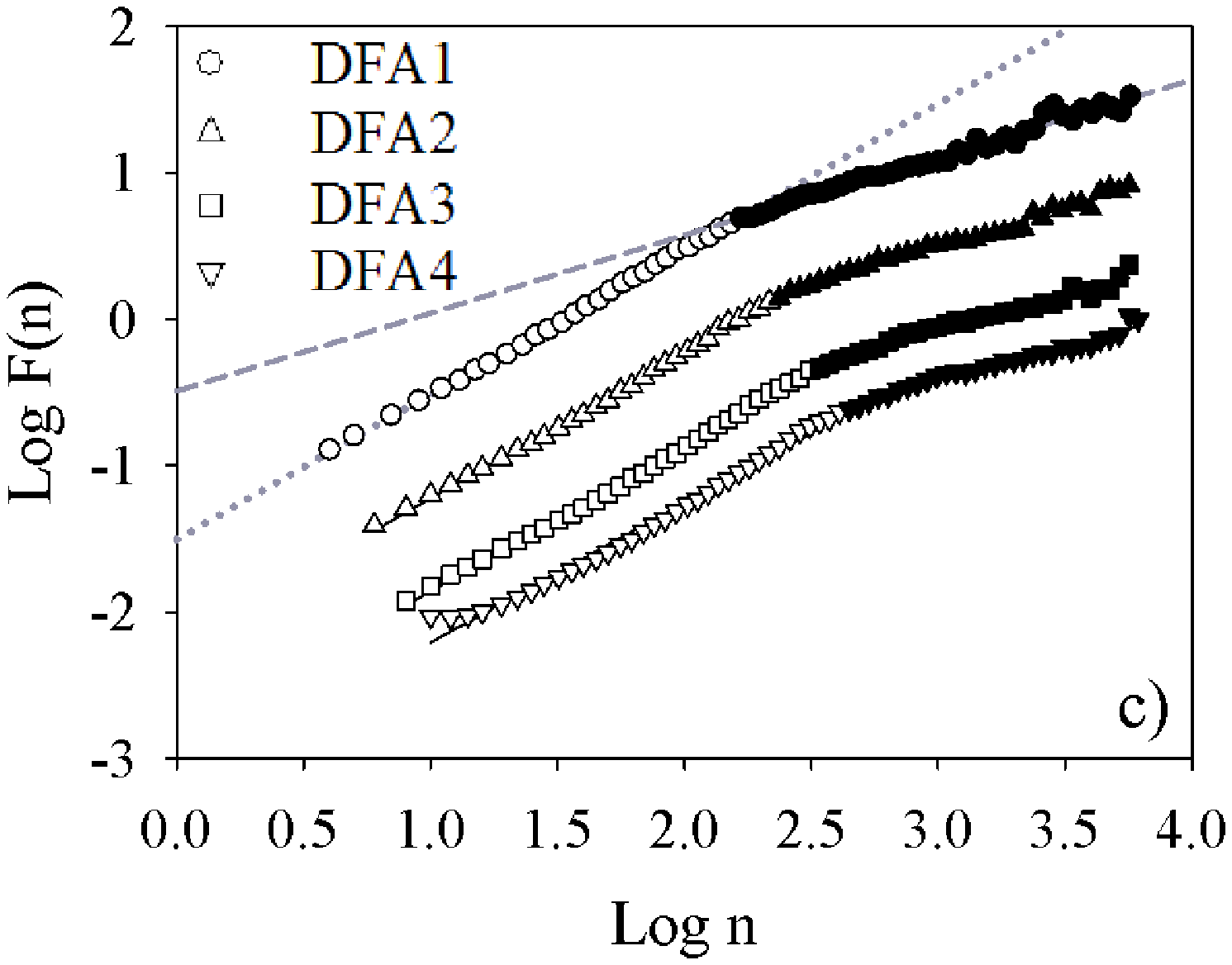}\\
\includegraphics[width=5cm]{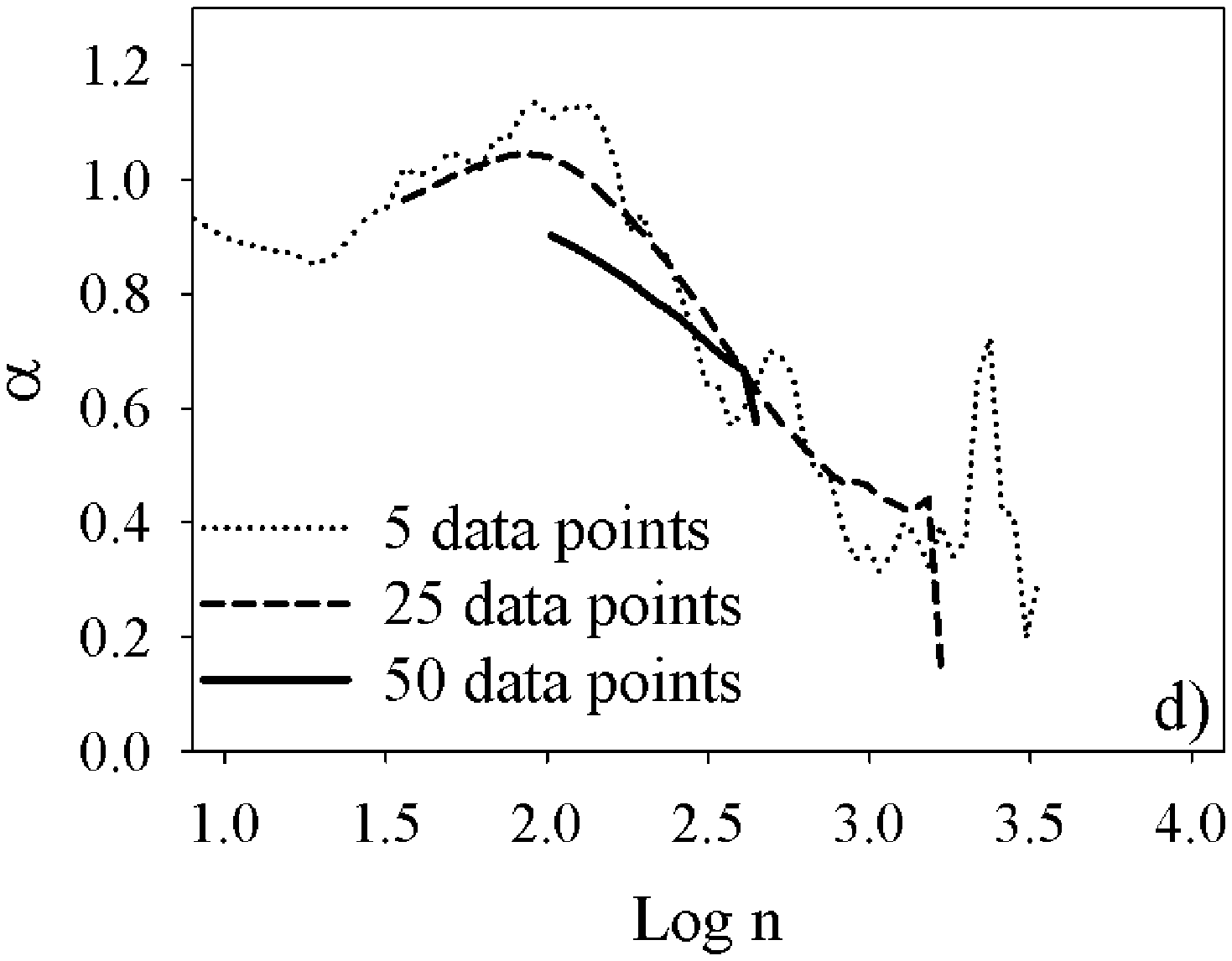}\includegraphics[width=5cm]{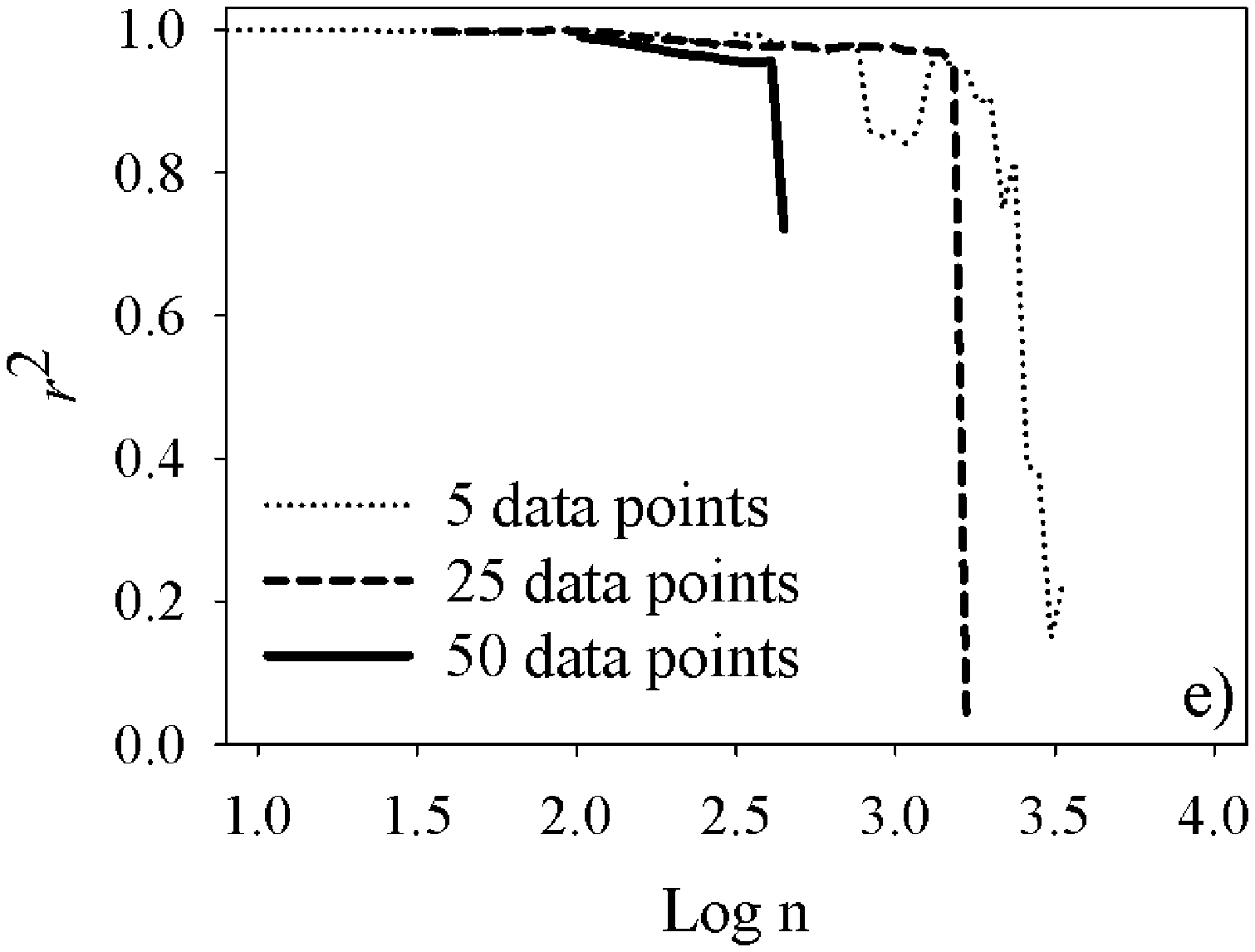}\includegraphics[width=5cm]{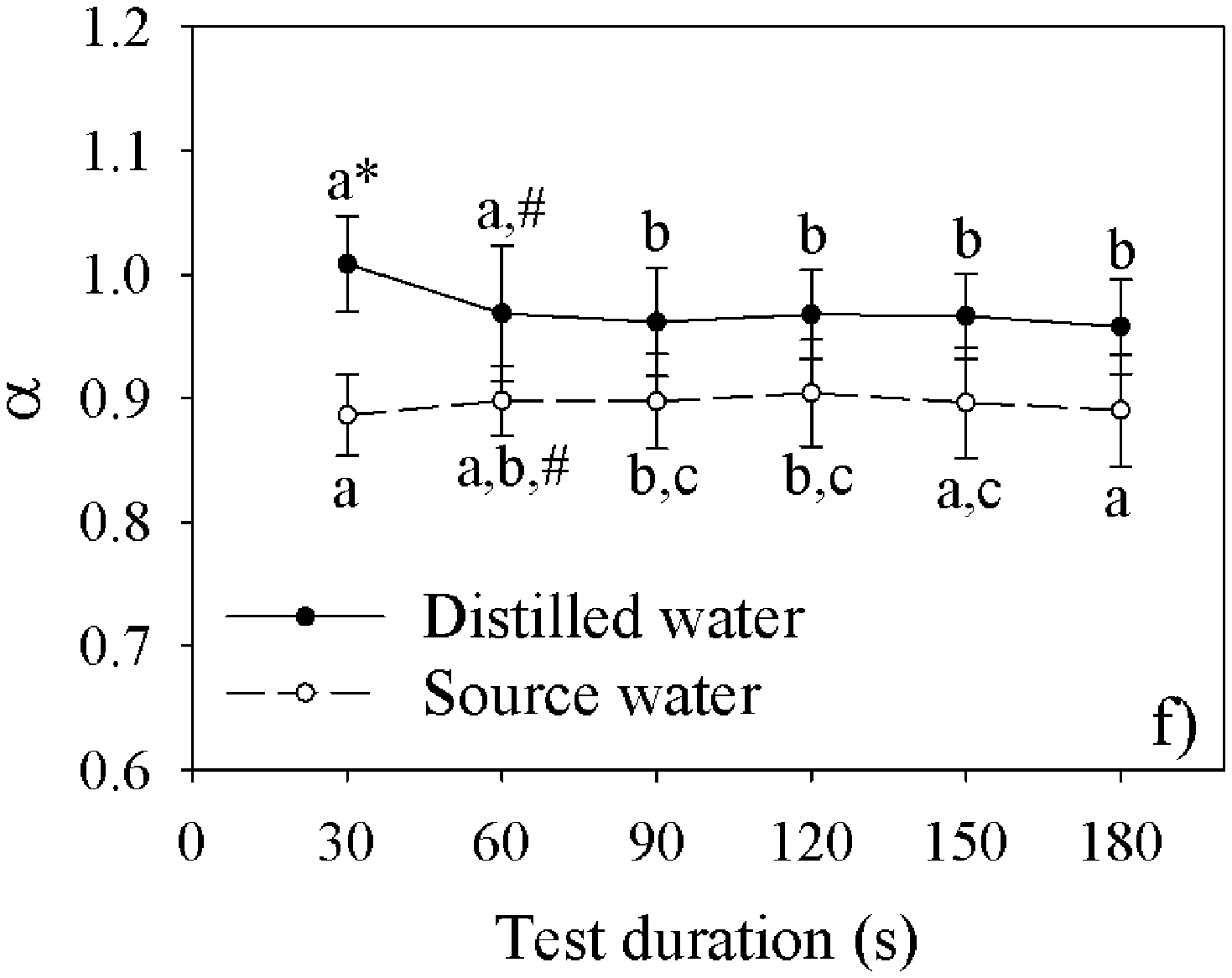}
\caption{Detrendend Fluctuation Analysis of locomotor temporal pattern of mosquito larvae Culex quinquefasciatus. a) Cumulative locomotor time series; b) Magnification of the same time series as "a", a similar pattern in observed for all levels of magnification. c) Example of trend elimination capability of DFA in time series using linear (DFA1), quadratic (DFA2), cubic (DFA3) and 4th order (DFA4) DFA. Observe that a clear change in slope is observed for all detrending orders. DFA3 was selected for group comparisons. Local slopes ($\alpha$) and e) $r^2$ of the corresponding linear fits estimated for different window sizes (5, 25 and 50 data points). Note the abrupt decrease in slope ($\alpha$) and $r^2$ for large windows, further indicating a crossover in the fluctuation function, $\alpha$ estimated with DFA, and f) for increasing test durations. Values are represented as mean $\pm$ S.E. A one-way repeated measures ANOVA was used to determine the effects water treatment (distilled or source water), and test duration (within-subject factor) as well as their interactions on the estimation of $\alpha$. A significant effect of test duration was observed ($P<0.05$). Test duration that do not share the same letter showed significant differences ($P<0.05$) in a LSD test.}
\end{figure}

\subsection{Effect of test duration}
The possible effect of test duration on the estimation of $\alpha$  in the locomotor data time series of quail and mosquito larvae was evaluated, by calculating $\alpha$ for increasing test durations. In Japanese quail time series a slight increase in the $\alpha$-value was observed for increases test duration, with an $\alpha \approx 0.79$ for 1 hour test duration that stabilized in $\alpha \approx 0.83$ for test durations $\geq 4$hs (Figure  1f). Although, this quantitative increase in $\alpha$ was observed, it does not represent a qualitative difference in scaling properties. In addition, throughout the day, no qualitative differences were observed in $\alpha$-values (range from 0.76 to 0.85) or on the $r^2$ of the linear fit ($r^2> 0.85$) when estimated for 1 hour time intervals throughout the day (data not shown), indicating the absence of a clear effect of circadian rhythms on the correlation properties of the locomotor pattern. Similar $\alpha$-values have been observed in previous studies of locomotor time series (with test durations $ \leq 1$ hour) in Japanese quail [18-20] and in a closely related species, the domestic chickens [9, 10], suggesting that the $\alpha$ -values within the range of 0.69 and 0.90 could be characteristic of the locomotor pattern of poultry. However, the exact $\alpha$-value depends on the experimental situation, such as the environment, the presence of stressors and age [9, 10, 18-20]. The stride interval of healthy adult human walking behavior (test durations $\leq 1$h) is also monofractal and was shown to have a higher $\alpha$ -value that varies between 0.84 and 1.1, remaining fairly constant despite substantial changes in walking velocity and mean stride intervals, but can be affected by age and disease [5, 34]. Also, the $\alpha$-values can vary substantially between behaviors in a particular species [7, 10, 12], for example in chicken the $\alpha$-value of vigilance behavior was 0.98 (S.D. = 0.042, range = [0.90-1.08]), while the walking pattern shows an $\alpha$-value of 0.70 (S.D. = 0.051, range = [0.58-0.81]) [10].

Figure 2f shows the effect of test duration on the estimation $\alpha_1$-values in mosquito larva. A significant difference between water treatments was observed for a test duration of 30min, this difference became less notorious for longer test durations approaching $\alpha_1\approx 0.95$. The $r^2$ of the estimation of $\alpha_1$ was always higher than 0.99. Similar $\alpha_1$ -values were observed in an independent study in larvae of {\sl Culex quinquefasciatus} in our laboratory [17], where $\alpha_1$-values were significantly decreased when larva were treated with lethal or sublethal doses of the essential oils of {\sl Lippia turbinata} and {\sl Lippia polystachya}. Recent studies have also shown that the $\alpha_1$-value could vary between mosquito larvae species depending on their ecological role in their habitat (Archilla, Kembro and Gleiser, unpublished data).

Before shorter time series can be used to evaluate the effects of a particular test treatment in the temporal pattern of a behavior in a given species, long-range correlations should be established previously. This is important because Maraun et al. [14] showed that a local slope larger than $\alpha=0.5$ for large scales does not necessarily imply long-memory. If the length of the time series is not sufficiently large compared to the time scales involved, also for short-memory processes $\alpha =0.5$ may not be reached. In their example, the empirical fluctuation function of a short-memory model of length $N=1000000 $ occurred at a constant slope ($\alpha=0.6$) between $l\log n \approx 2.6$ ($n\approx 600$) and $\log n \approx 3.8$ ($n \approx 6000$). Only on larger scales the slop reduces to $\alpha=0.5$. This process in which the fluctuation function converges to $\alpha=0.5$ cannot be observed for shorter time series ($N= 70492$).

In all, our results show that although test duration can slightly influence the estimation of $\alpha$, no qualitative differences were observed between different test durations that varied between 1 and 12 h. Also, considering the consistency observed between different experiments and laboratories in the values of $\alpha$, certain ranges of $\alpha$-value and autocorrelation structure (mono or multifractality) could be characteristic of a certain behavior of a given species. Therefore, once long-range correlations are established (see previous sections) using ``long" time series, shorter time series could be used during testing, in order to evaluate, for example, changes in $\alpha$-value due to a particular treatments, such as modifications in the environment, stressors or neuroactive substances.

\section{ Frequency distribution of the duration of immobility and mobility events (FDD-I and FDD-M, respectively)}

\begin{figure}[h]
\centering
\includegraphics[width=5cm]{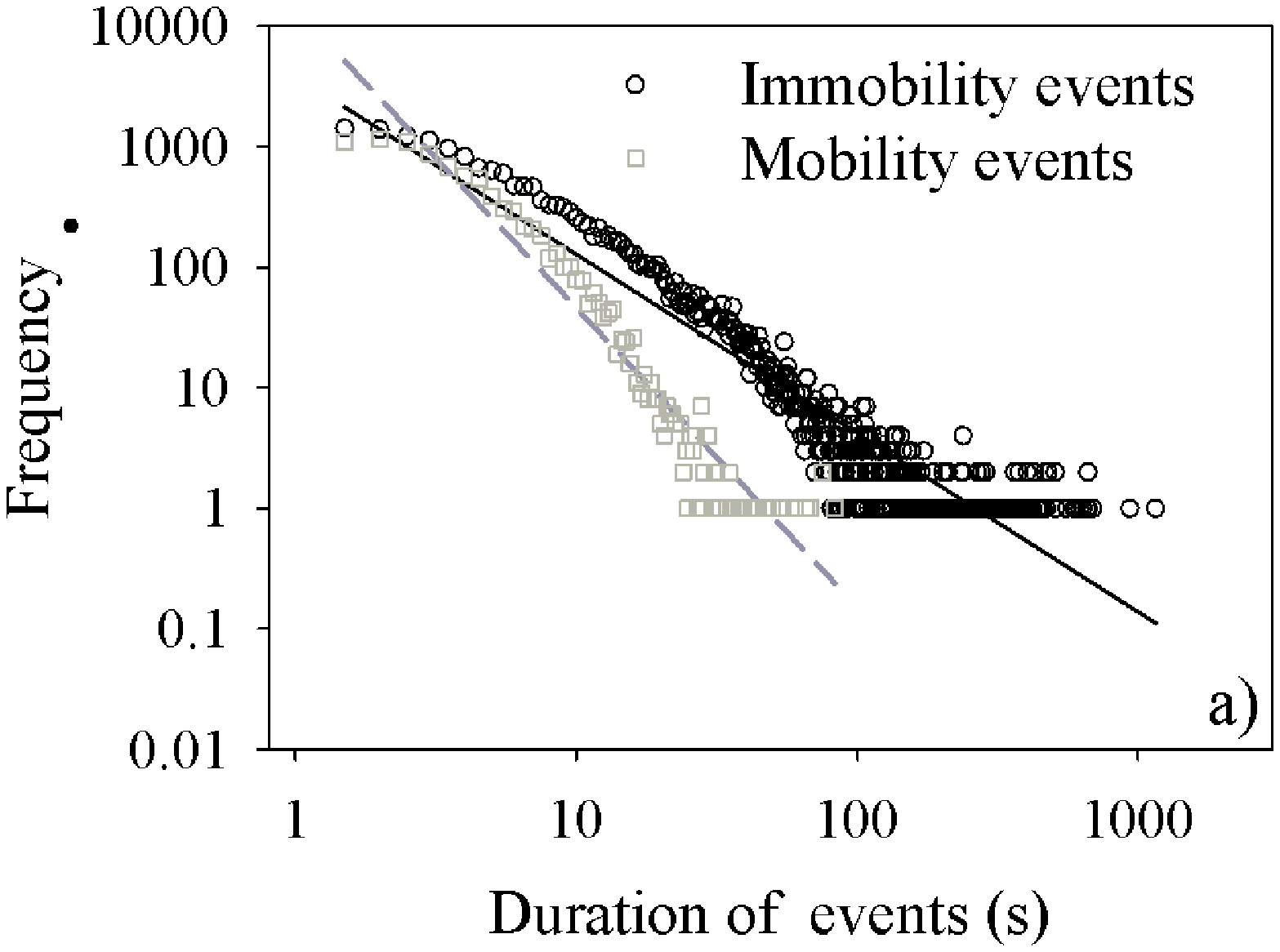}\includegraphics[width=5cm]{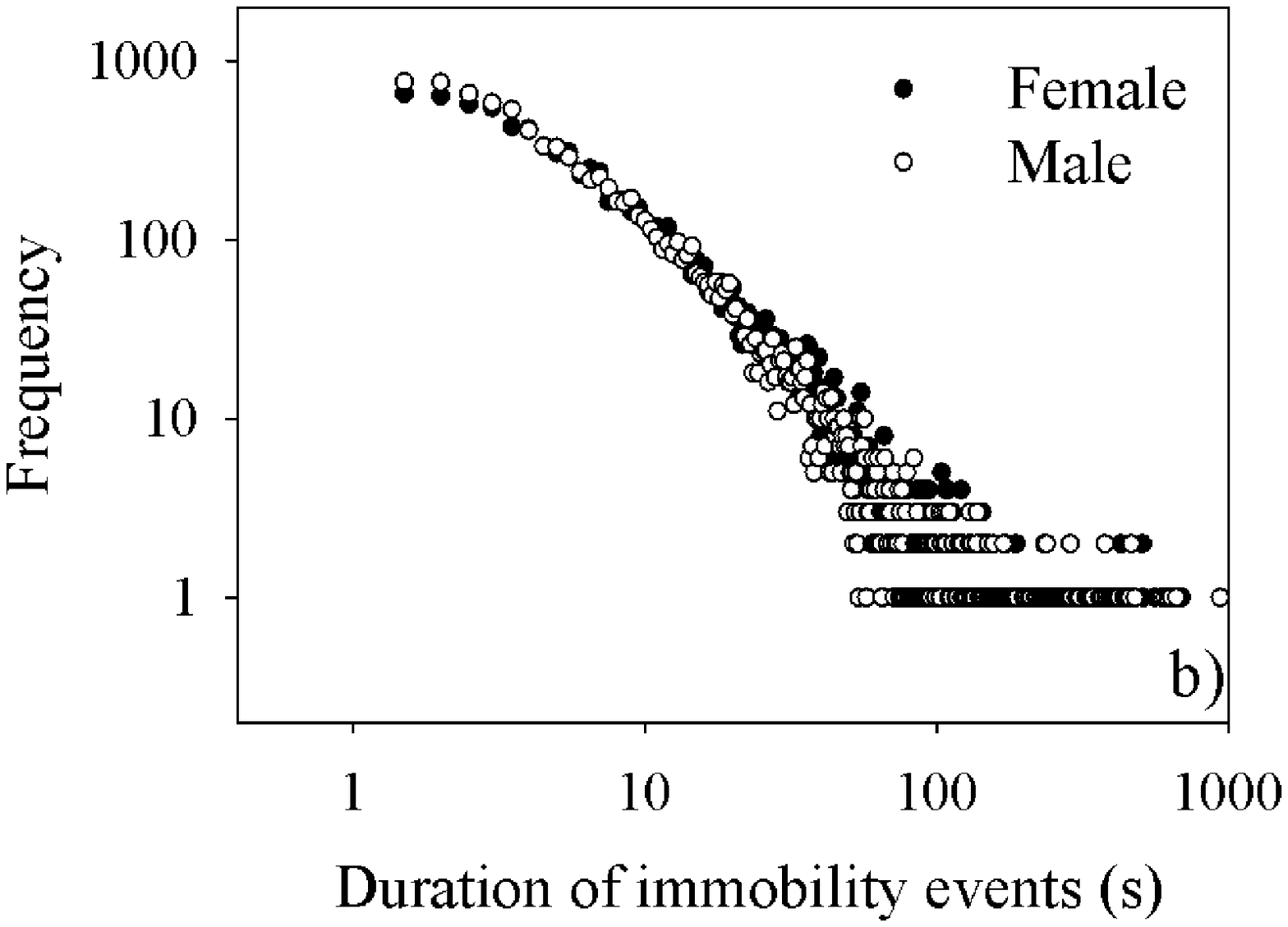}\includegraphics[width=5cm]{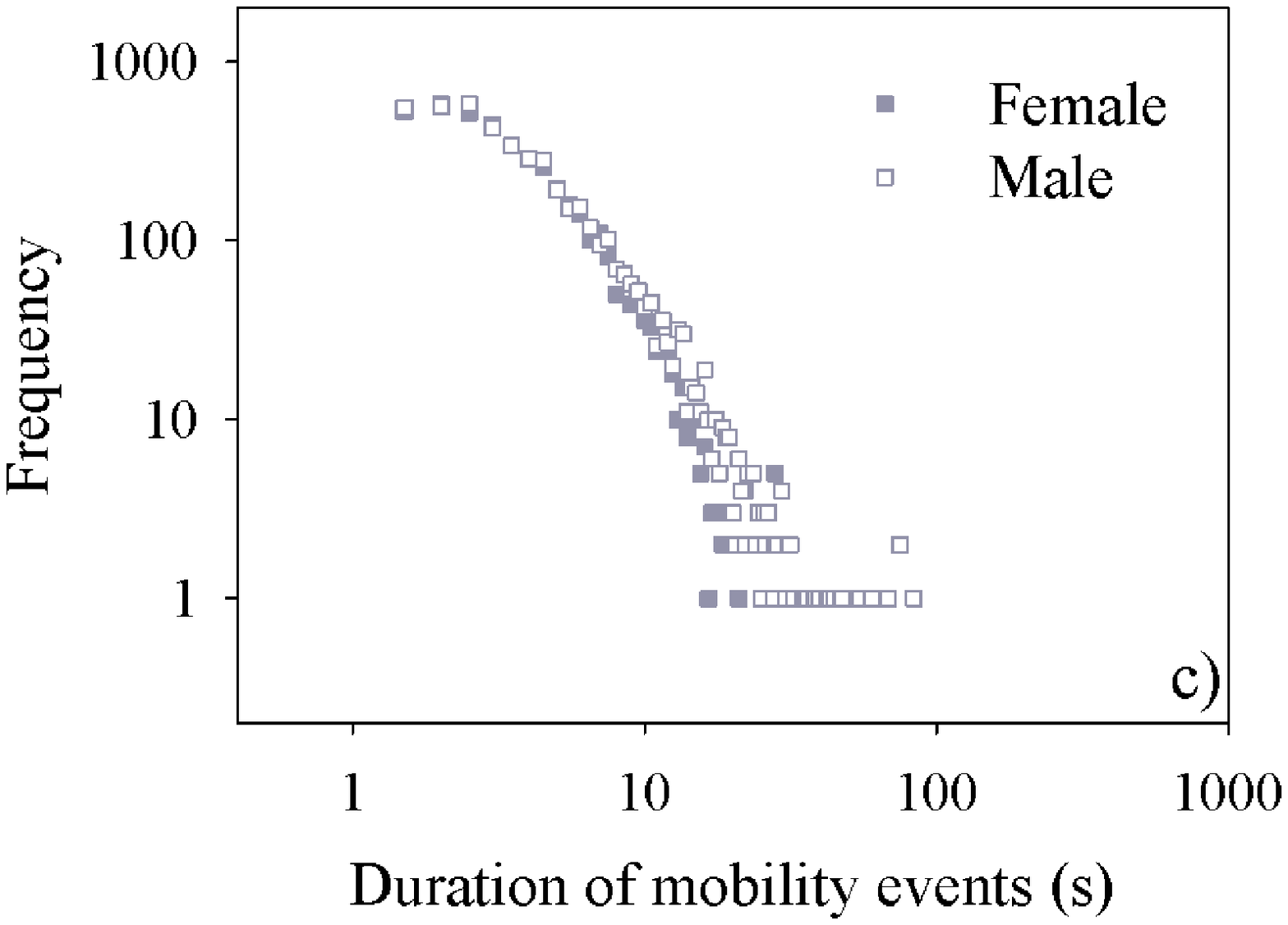}
\caption{Frequency distribution of the duration of immobility and mobility events of Japanese quail calculated with: a) data from all animals is combined, and b-c) with data discriminated taking into consideration sex. Females were represented by filled symbols, while males were shown as open symbols.}
\end{figure}

The FDD-I and FDD-M was analyzed by plotting the frequency vs. the duration of immobility or mobility events, respectively, using a double logarithm scale. We evaluated whether the frequency distribution adjusted to a linear fit in a log-log plot when all animal data (Fig 3a and 4a) and when data from each treatment was pooled together (female/male (Fig 3b-c) or distilled/source water (Fig 4b-c) are plotted separately). Also we evaluated the FDD-I and FDD-M of each animal separately in order to allow statistical comparisons between groups.  When a linear fit was achieved, it was considered as indicative of a power-law (fractal) distribution. The slope of the linear fit is known as the scaling factor (S) and was determined for both the immobility ($S_I$) and the mobility ($S_M$) events. We also looked at the $r^2$ value of the distribution when plotted on a semi-log scale, which would correspond to an exponential distribution. In order to consider the frequency distribution a power law, the $r^2$ value must be higher in the fit to the log-log plot than the semi-log plot.

In Japanese quail, many long duration of immobility and mobility events were present, indicating an asymmetric frequency distribution of the duration of the immobility or mobility events. In figure 3a these frequency distributions show linear fits in a log-log plot for immobility events $S_I= -1.48\; (r^2 = 0.86)$, and $S_M = -2.48\; (r^2 = 0.92)$ when the data for all animals is combined. Similar values (Figure  3b,c) were found when data was discriminated taking into consideration sex (Female: $S_I= -1.38\; (r^2 = 0.85)$, $S_M=-2.40 \;(r^2 = 0.90)$, and Male: $S_I=-1.33\; (r^2 = 0.83)$, $S_M = -2.23 \;(r^2 = 0.92)$). When inter-individual variations were evaluated at different test durations, the mean value of $S_I$ (Figure  5c) and $S_M$ (Figure  5g) for the female and male groups became more negative with larger values of $r^2$ (Figure  5e,i) for longer test durations approaching the values estimated when the data from all the individuals of each group was combined. Interestingly, these results cannot be explained due to fluctuations in the time mobile or duration of immobility and mobility events throughout the day (Figures 5a,b,d,f,h,j). Although the amount of data obtained per animal during the one hour trail is not enough to accurately represent the tails of the numerical distributions ($0.58 > r^2 > 0.79$; Figure  5f,j); nevertheless, Figure 5d shows that the mean value of $S_I$ fluctuates throughout the day between -0.37 and -0.77, and becomes only slightly more negative towards the end of the day. Similar values (mean values of $S_I= 0.66 \pm 0.05$) were observed in an independent 40 min study also in Japanese quail [19]. Also, the mean value of $S_M$ remains fairly constant throughout the day, ranging between -0.79 and -1.03 (Figure  5h).
\begin{figure}[h]
\centering
\includegraphics[width=5cm]{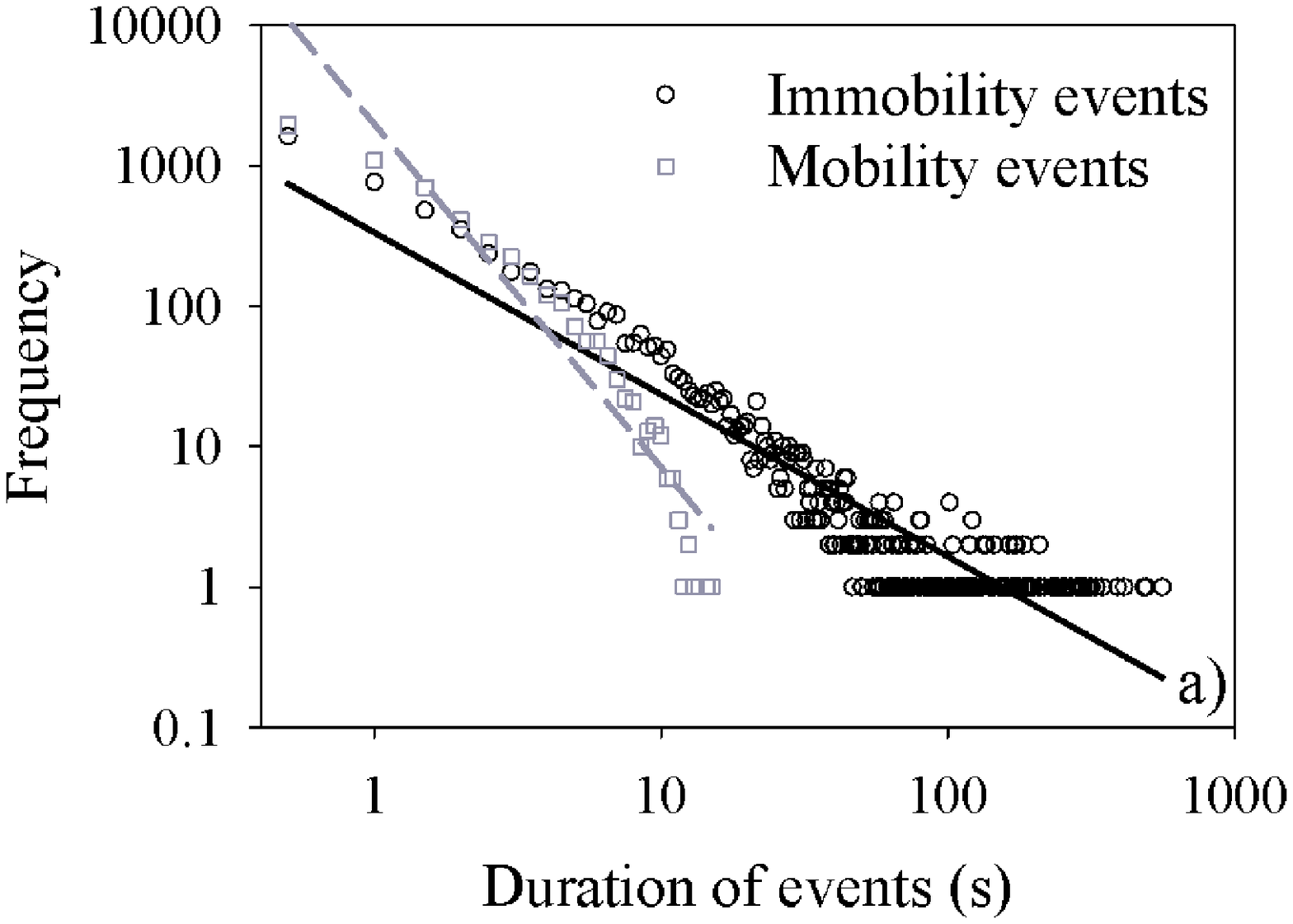}\includegraphics[width=5cm]{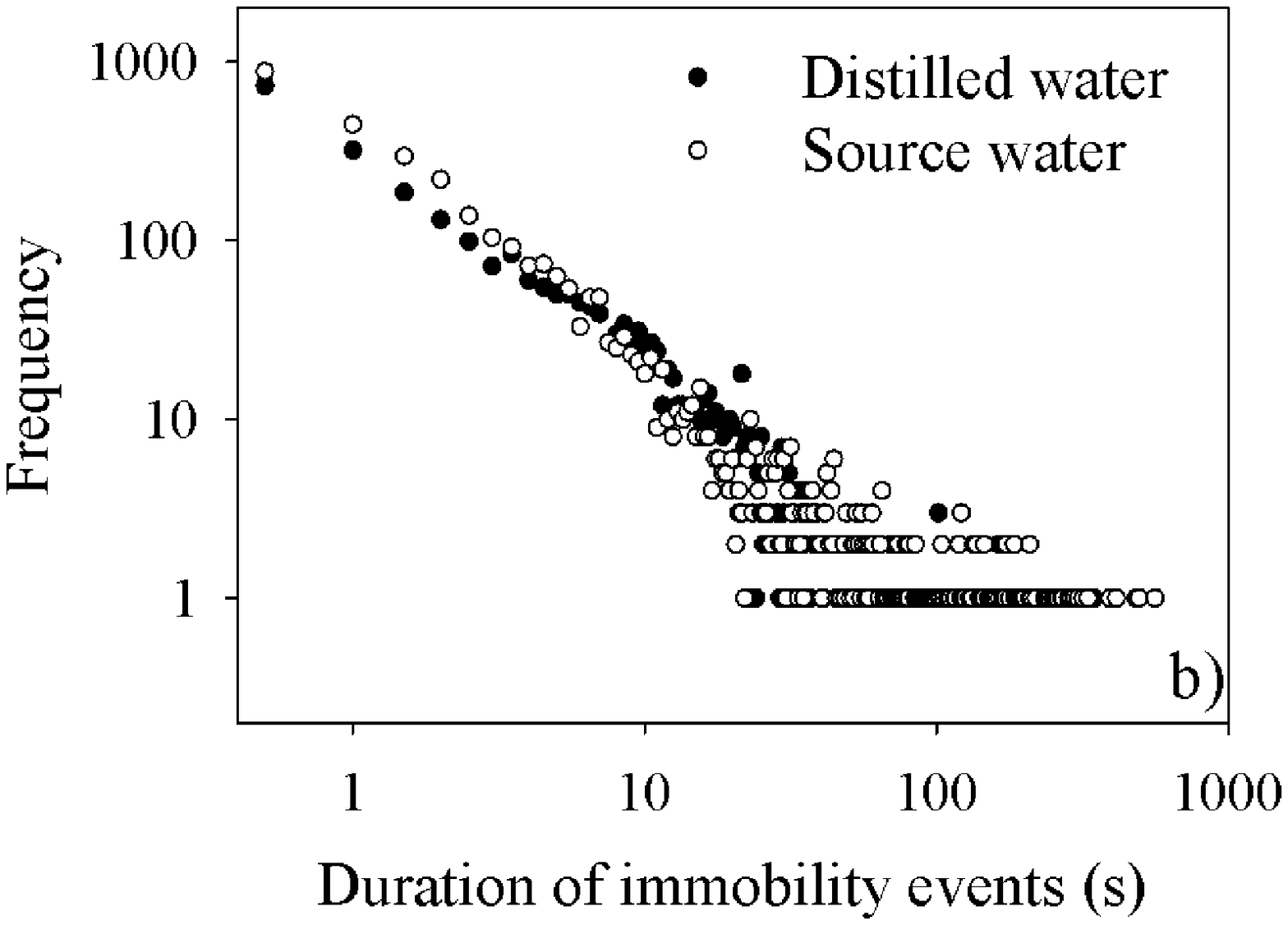}\includegraphics[width=5cm]{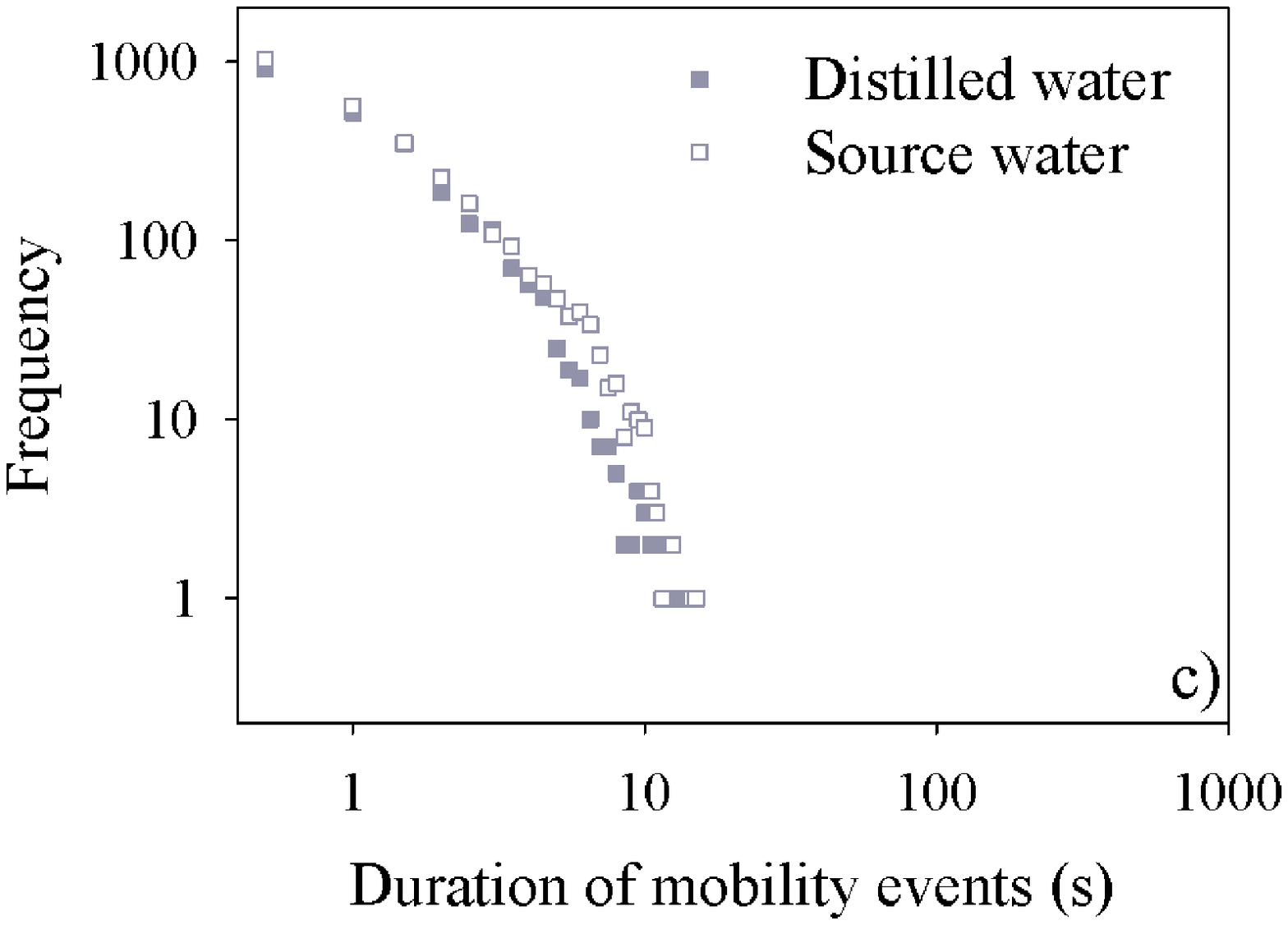}
\caption{Frequency distribution of the duration of immobility and mobility events of mosquito larve {\sl Culex quinquefasciatus} calculated with: a) data from all animals is combined, and b-c) with data discriminated taking into consideration water treatment. Source water treatment was represented by filled symbols, while distilled water treatment was shown as open symbols.}
\end{figure}

The majority of the studies that evaluate the distribution of the duration of behavioral events combine the data of many animals to obtain the frequency histograms, especially if the test has a short duration [13, 26, 35, 36]. In this study when the data from the one-hour records are combined, similar values were obtained of $S_I= - 1.03 \;(r^2 = 0.81)$; (Female: $S_I = -0.81\;(r^2 = 0.75)$; Male: $S_I=-0.97 \;(r^2 = 0.77)$), and $S_M = -1.56\; (r^2 = 0.86)$ (Female: $S_M= -1.27 \;(r^2 =0.76)$ and Male: $S_M=-1.51 \;(r^2 = 0.80)$). These values are a substantial improvement on the estimation of these parameters for short test duration (1 h), showing values closer to those estimated when the test lasted more than 4hs. However, the combination of data from many animals result in the loss of valuable information regarding inter-individual variations, which are of importance in order to characterize population variability and to study differences between treatment groups. For example, quail stimulated to explore for food (food was scattered on the floor of the home box after 3 hours of feeder withdrawal) show a significantly more negative $S_I$ scaling factor, and a tendency toward a more negative $S_M$ scaling factor, than their un-stimulated counterparts [19]. In all, our results show that when evaluating the frequency distribution of the duration of immobility and ambulation events it is important to use long time series for estimation of $S_I$ and $S_M$, in order to improve estimation of parameters and to improve the evaluation of inter-individual variability in the population.

In mosquito larvae, asymmetric frequency distributions of the duration of the immobility or mobility events were also observed. In Figure 4a these frequency distributions show linear fit in a log-log plot of all treatments for immobility events $S_I = -1.15\; (r^2 = 0.83)$, and $S_M = -2.44\; (r^2 = 0.88)$. Similar values (Figure  4b,c) were found when data was discriminated taking into consideration water treatment (Distilled: $S_I=-1.12 \;(r^2  = 0.81)$, $S_M = -2.36\; (r^2 = 0.93)$ , and Source: $S_I = -0.96 \;(r^2 = 0.79)$, $S_M=-2,15\; (r^2 = 0.89)$).

In mosquito larva $S_M$ also can be fitted to an exponential (Total data: $slope= -0.22\; (r^2 = 0.97$); Distilled water: $slope = -0.23\; (r^2 = 0.95$); Source water: $slope=-0.20 \;(r^2 = 0.96$)). This contrasted our previous results in quail where exponential fits have $r^2 \leq 0.64$ for $S_M$, and in both species $S_I$ fit an exponential with an $r^2\leq  0.32$. When inter-individual variations were evaluated at different test durations, the mean value of $S_I$ (Figure  6c) and $S_M$ (Figure  6e) for the both water treatment groups became more negative with larger values of $r^2$ (Figure  6d,f) for longer test durations approaching the values estimated when the data from all the individuals of each group were combined. As in Japanese quail, these results cannot be explained due to fluctuations in the time mobile (Figure  6a,b) or duration of immobility and mobility events throughout the day (data not shown).

Anteneodo and Chialvo [35] showed in rat, evaluated in their home-cage during
9 consecutive days, that from few seconds to several thousand seconds (about 1 hour), the
distribution of inter-event times (equivalent to immobility) decays as a power law (with
scaling exponent falling within the interval of  $1.75 \pm 0.05$) for all six animals. In contrast, the distribution of the duration of motion (mobility) episodes did not possess a characteristic time scale, and was described by a superposition of two exponentials with characteristic times of the order of 1 and 4 s, close to the smaller data resolution and to the average duration of motion episodes, respectively.

In our study, quail and mosquito larvae both show frequency distributions of immobility and mobility events that decay as a power law. In addition, the value of the scaling exponent for immobility events ($S_I$) were lower in quail and in mosquito larva (-1.48 and -1.15, respectively) compared to the values observed by Anteneodo and Chialvo [35]. In both species the scaling exponent of mobility events $S_M$ (-2.48 in quail and -2.44 in mosquito larvae) was larger than $S_I$ indicating that mobility events are frequently of shorter duration than immobility events.

Our results show that in contrast with DFA (where no qualitative effect of test duration was observed on the estimation of the $\alpha$-value) when evaluating the frequency distribution of the duration of immobility and ambulation events it is important to use long time series for estimation of $S_I$ and $S_M$, in order to improve estimation of parameters. For short time series ($<10000$ data points) pooling data from animals is a way to improve estimation of scaling parameters, but eliminates the possibility of evaluating variability in the population, which is very important in animal behavior studies due to the natural large variation in a behavioral response even between animals of a same species, sex, age, and reared identically.

\begin{figure}[h]
\centering
\includegraphics[width=5cm]{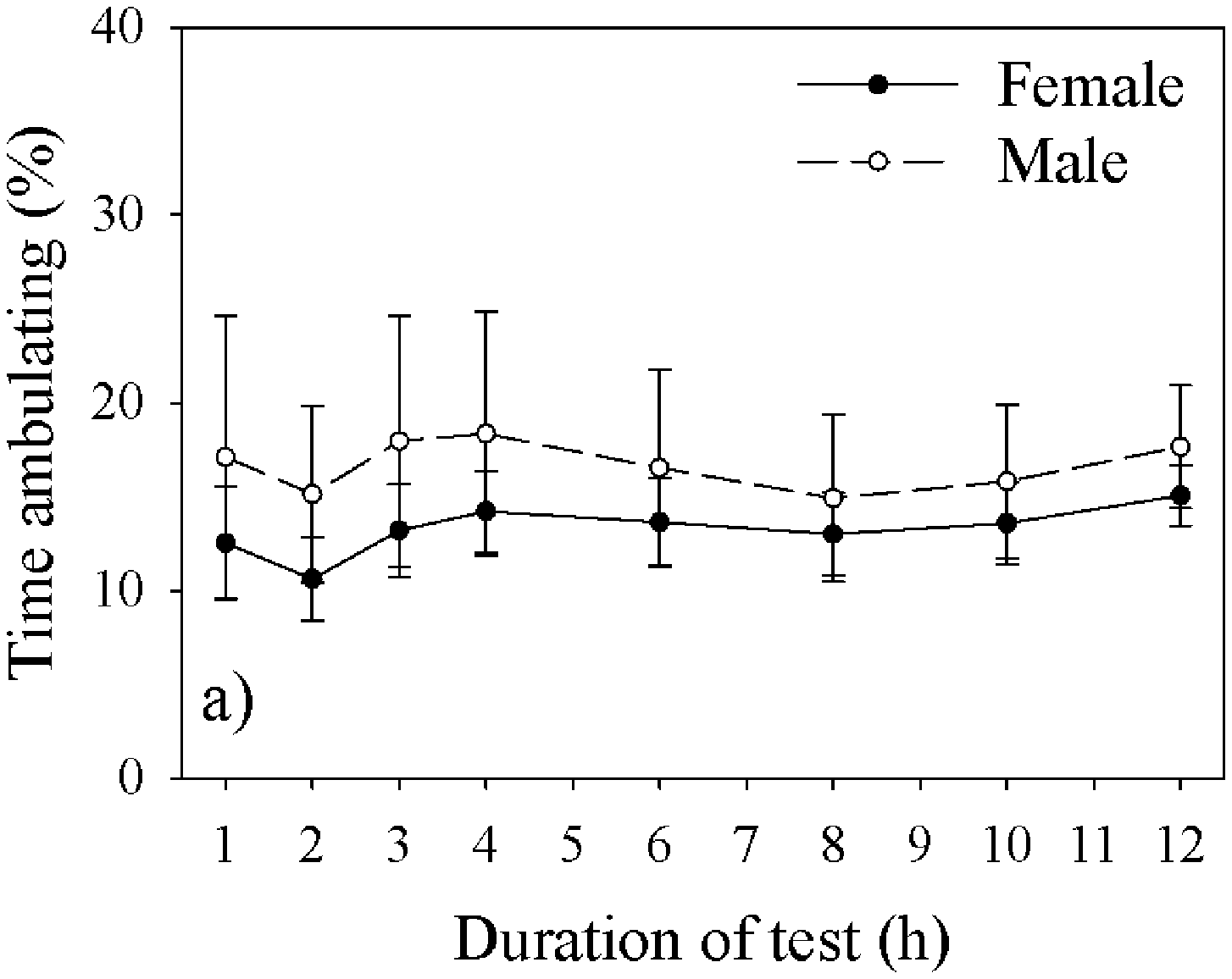}\includegraphics[width=5cm]{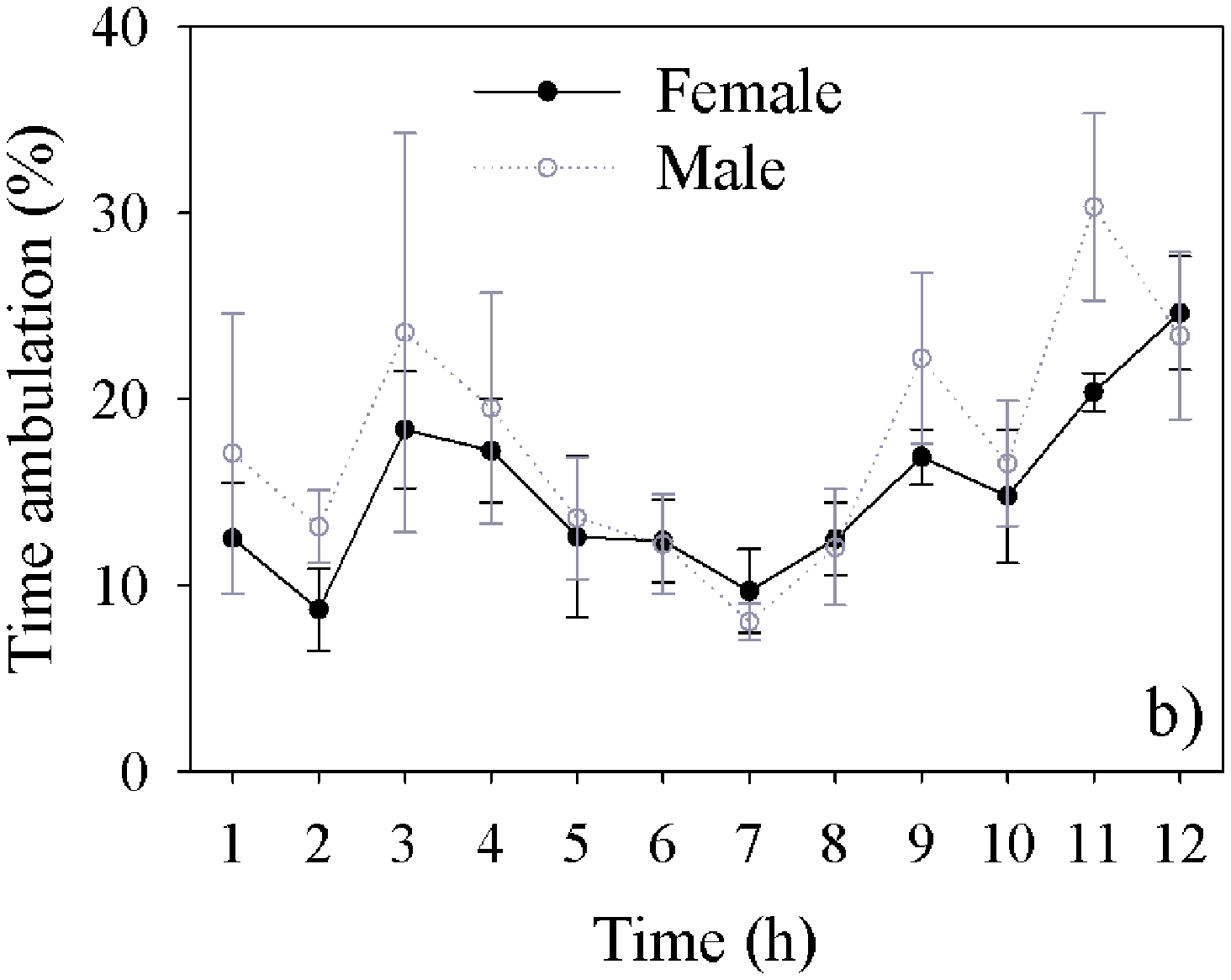}\includegraphics[width=5cm]{Kembro_Fig5b}\\
\includegraphics[width=5cm]{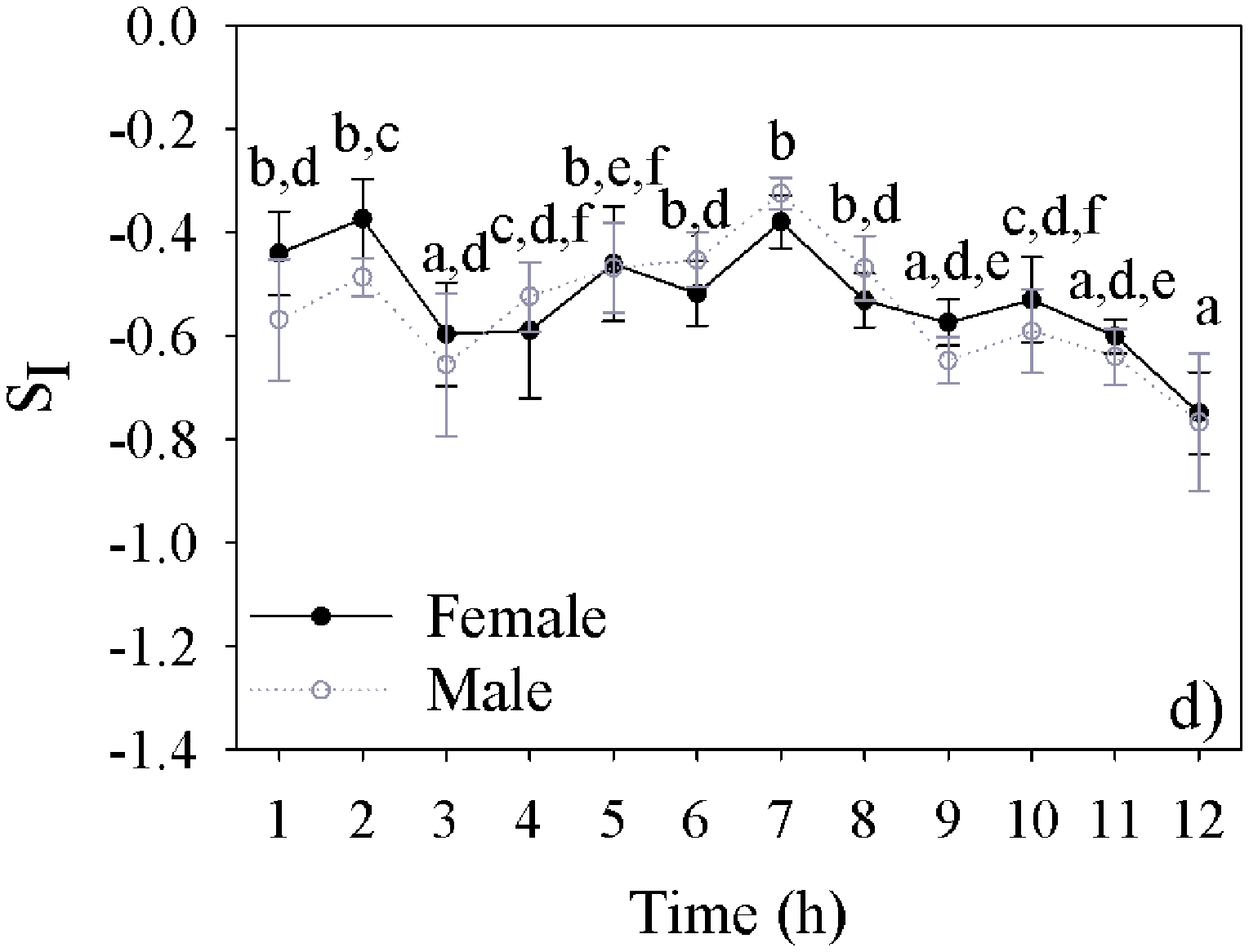}\includegraphics[width=5cm]{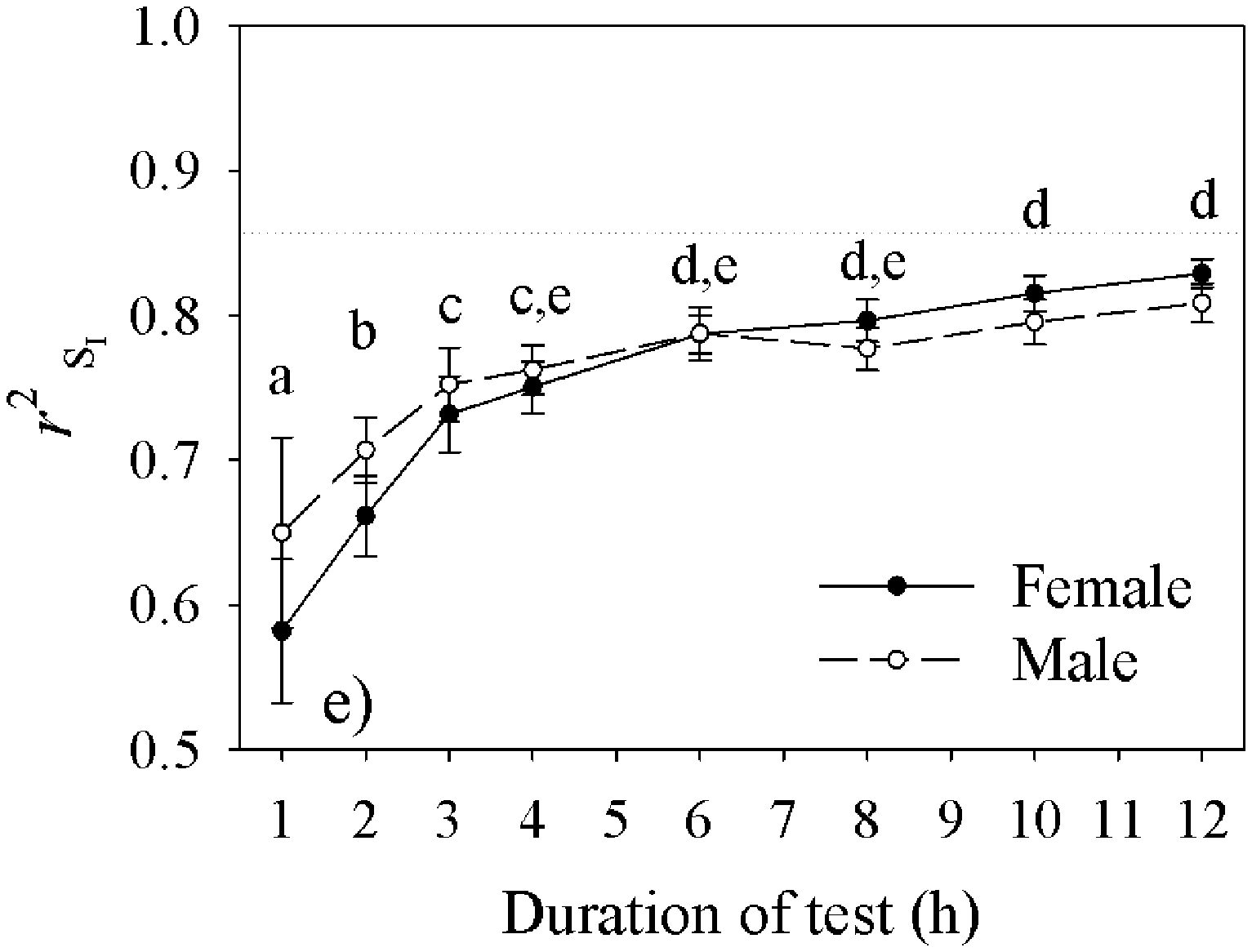}\includegraphics[width=5cm]{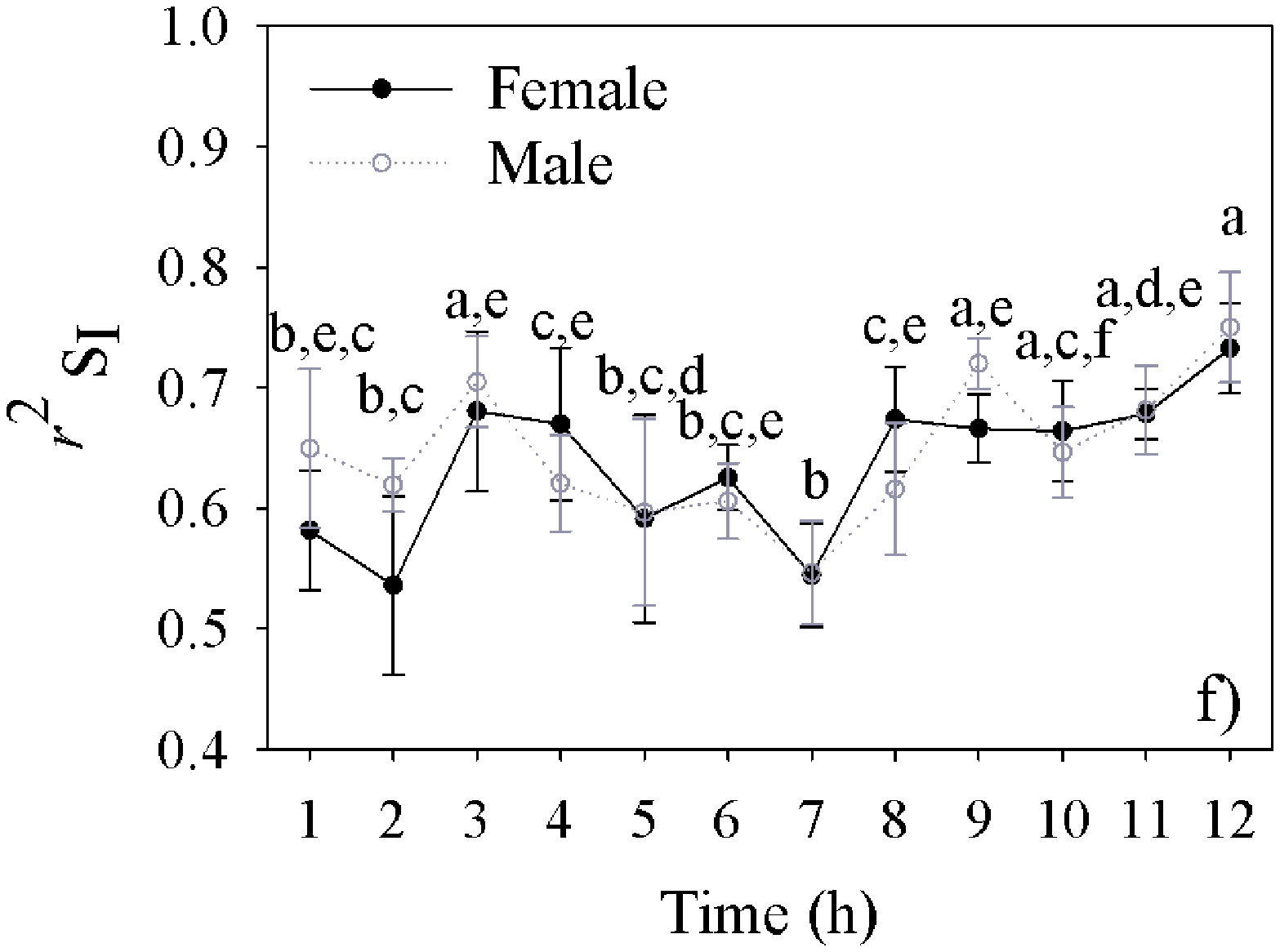}\\
\includegraphics[width=5cm]{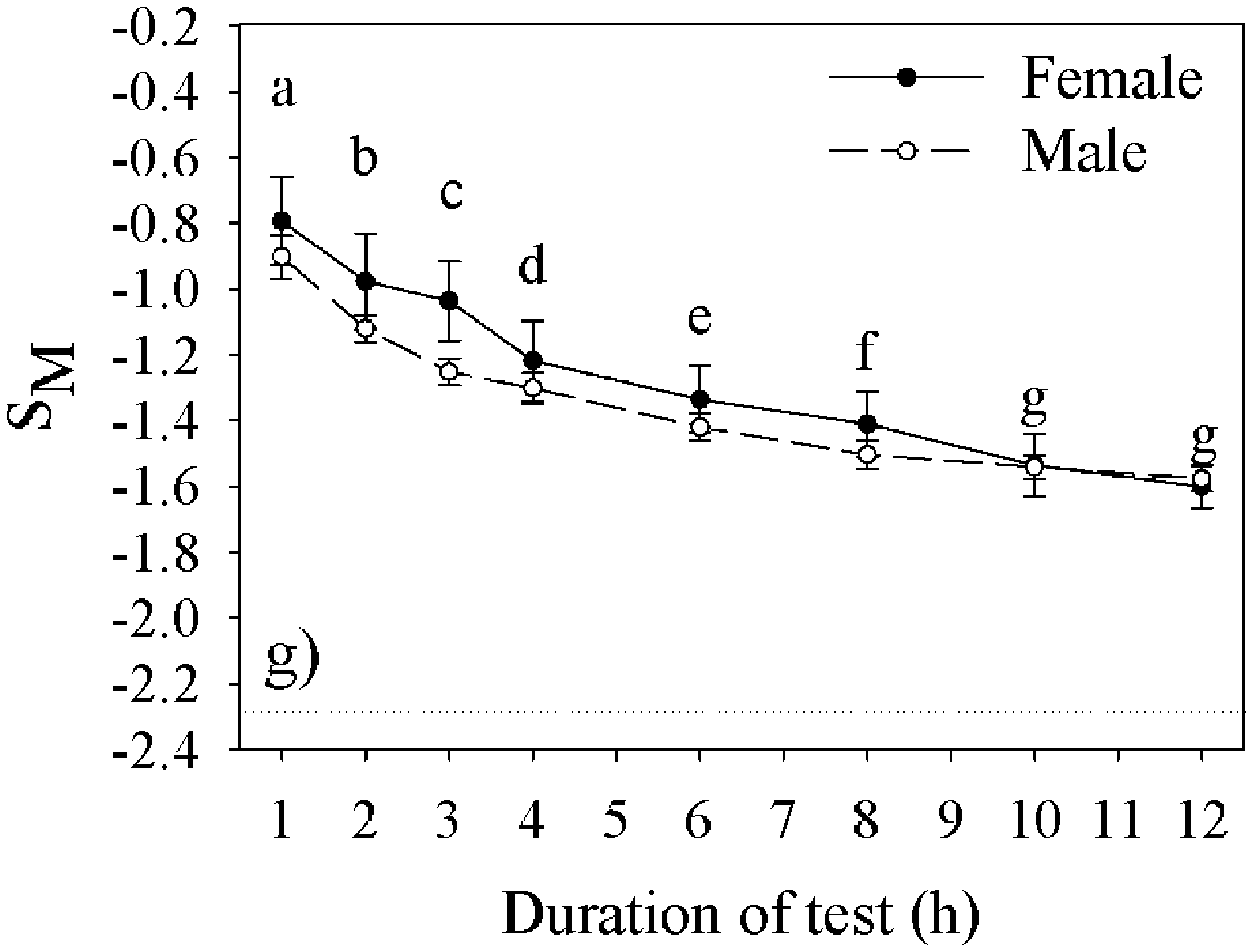}\includegraphics[width=5cm]{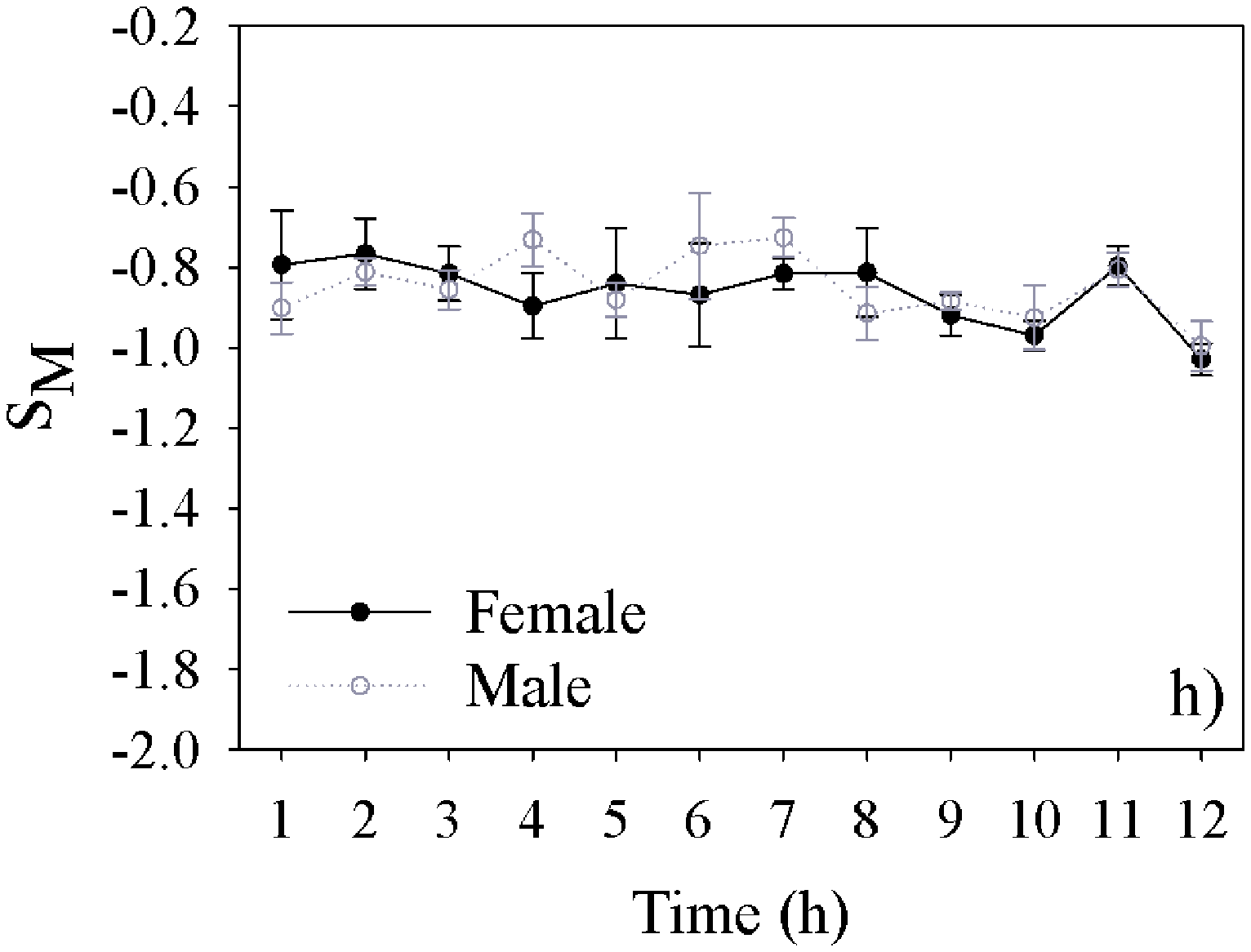}\includegraphics[width=5cm]{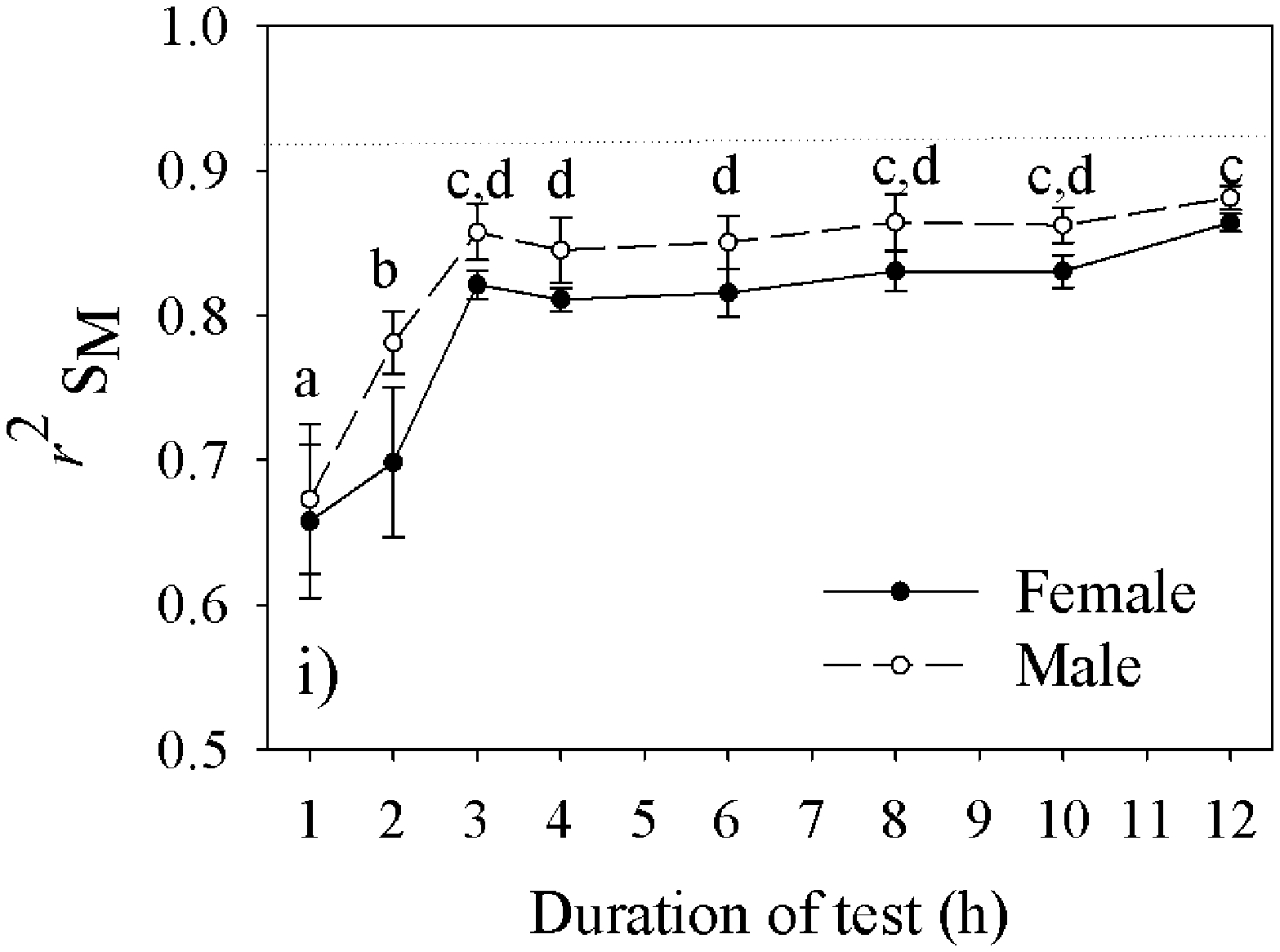}\\
\caption{Parameters of locomotor activity in female and male Japanese quail. a-b) Percentage of time ambulating, Scaling exponents calculated as the slop of the log-log plot of frequency vs. duration of the immobility or ambulation events, c-d) $S_I$ and g-h) $S_A$, respectively, and e-f, i-j) the respective $r^2$ of the linear fit estimated for: a,c,e,g,i) increasing test durations and b,d,f,h,j) at one hour intervals throughout the day. Values are represented as mean $\pm$ S.E. A one-way repeated measures ANOVA was used to determine the effects gender (female and male), and the test duration (within-subject factor) as well as their interactions on the estimation of variable. When a significant effect of test duration was observed (P$<0.05$) a LSD test was performed. Test duration that do not share the same letter showed significant differences ($P<0.05$).}
\end{figure}

\begin{figure}
\centering
\includegraphics[width=5cm]{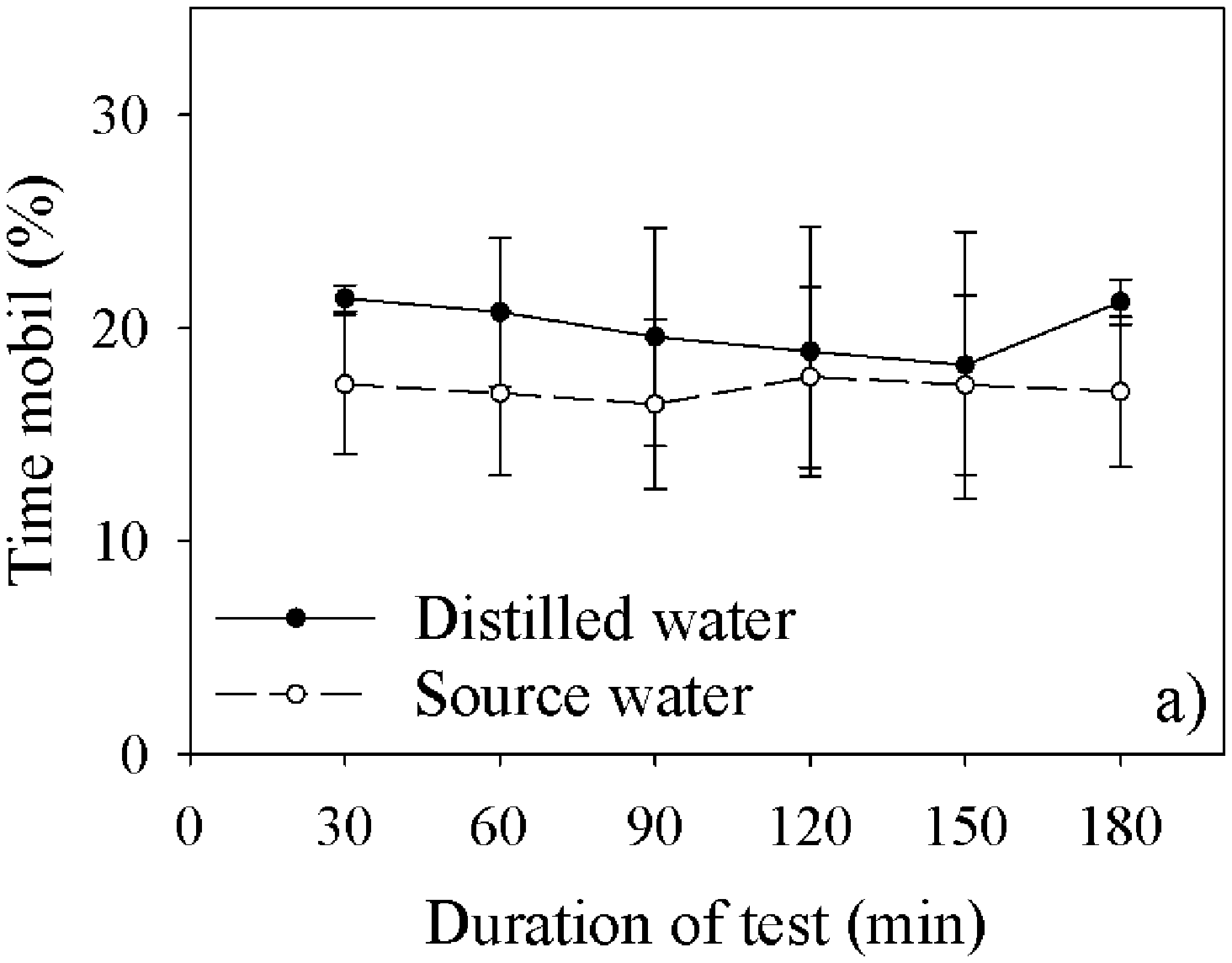}\includegraphics[width=5cm]{Kembro_Fig5b}\includegraphics[width=5cm]{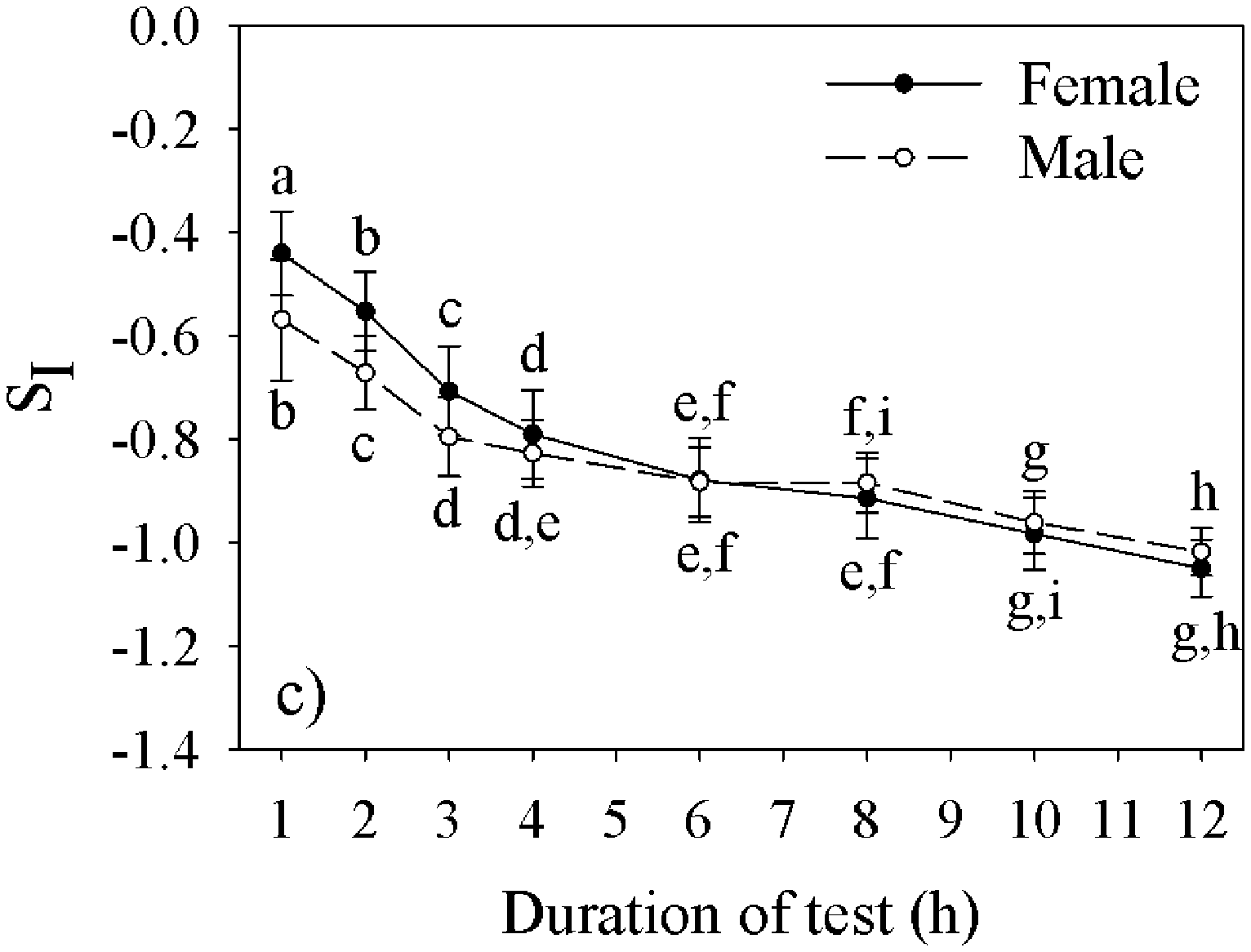}\\
\includegraphics[width=5cm]{Kembro_Fig5d}\includegraphics[width=5cm]{Kembro_Fig5e}\includegraphics[width=5cm]{Kembro_Fig5f}\\
\caption{Parameters of locomotor activity of mosquito larvae of Culex quinquefasciatus. a-b) Percentage of time ambulating, Scaling exponents calculated as the slop of the log-log plot of frequency vs. duration of the immobility or ambulation events, c-d) $S_I$ and g-h) $S_A$, respectively, and e-f, i-j) the respective $r^2$ of the linear fit estimated for: a,c,e,g,i) increasing test durations and b,d,f,h,j) at one hour intervals throughout the day. Values are represented as mean $\pm$ S.E. b) A one-way repeated measures ANOVA was used to determine the effects water treatment (distilled or source water), and test duration (within-subject factor) as well as their interactions on the estimation of the variable. When a significant effect of test duration was observed ($P<0.05$) a LSD test was performed. Test duration that do not share the same letter showed significant differences ($P<0.05$).}
\end{figure}

\section{Summary}

The goal of this paper was to evaluate the presence of long-range correlation in animal behavior time series, in particular the temporal pattern of locomotor activity of Japanese quail ({\sl Coturnix coturnix}) and mosquito larva ({\sl Culex quinquefasciatus}), using DFA.  In our study, we discuss the following points: 1) the establishment by hypothesis testing of the absence of short-term correlation, 2) the accuracy of line regression estimation in the log-log plot of the fluctuation function, 3) the elimination of artificial crossovers in the fluctuation function, and 4) the influence of length of the time series in the accuracy of estimation.
These aspects have been previously studied using different systems such as statistical model systems [14,22,23,39,40],  temperature records  [14], DNA sequences  [30, 37], cardiac RR interval time series [39] or in financial economics [38],  but to our knowledge this is the first time these aspects have been systematically studied directly in animal behavior experiments.

In this study, we observed by hypothesis testing that the existence of long-range correlations for all series was statistically significant at level 5\%. We also studied artificial crossover with several detrending order.

Many authors have stated that the length of the time series is the most alarming factor when determining accuracy in scaling properties estimation. Short time series can suggest long-range correlation just because the series has not reached its plateau. Thus, many authors have suggested the use of DFA only with long series. In our case, the Japanese quail ({\sl Coturnix coturnix}) behavior can be monitored during relatively long periods of time without any significant change in experimental conditions. Mosquito larva ({\sl Culex quinquefasciatus}) changes it´s behavior after a significant number of hours, so test duration is mandatory restricted.

We also compared DFA to another fractal analysis, frequency distribution of the duration of behavioral events. Our results show that when using this frequency statistic, long time series are very important for obtaining high accuracy in the estimation of $S_I$ and $S_M$. With DFA, there was no qualitative effect of test duration when evaluating the frequency distribution of the duration of immobility and ambulation events.

\section{Acknowledgements}

This research was supported by grants from FONCYT, SECyT UNC and CONICET, Argentina. JMK, AGF, RG and RHM are career members of CONICET, Argentina. JMK is corresponding author.

\end{document}